\documentclass[11pt]{article}

\usepackage[margin=1in]{geometry}
\usepackage{setspace}
\usepackage{times}
\usepackage{microtype}

\usepackage{titling}
\usepackage{authblk}
\usepackage{abstract}
\usepackage{amsmath, amssymb, amsfonts, bm}
\usepackage{graphicx}
\usepackage{booktabs}
\usepackage{multirow}
\usepackage{array}
\usepackage{algorithm}
\usepackage{algpseudocode}
\usepackage{tikz}
\usetikzlibrary{arrows.meta, positioning, fit, backgrounds, calc, shapes.geometric}

\usepackage{endnotes}
\usepackage{xurl}
\usepackage{xcolor}

\usepackage[round,authoryear]{natbib}
\bibpunct[, ]{(}{)}{,}{a}{}{,}%
\def\bibfont{\small}%
\usepackage[colorlinks=true, linkcolor=blue, citecolor=blue, urlcolor=blue]{hyperref}

\definecolor{GenRed}{RGB}{228, 26, 28}
\definecolor{GenBlue}{RGB}{55, 126, 184}
\definecolor{GenPurple}{RGB}{152, 78, 163}
\definecolor{GenGray}{RGB}{153, 153, 153}

\newcommand{\gmark}[1]{\textcolor{#1}{\rule{1.2ex}{1.2ex}}\hspace{0.5ex}}
\newcommand{\up}{}
\newcommand{\down}{}
\newcommand{\TABLE}[3]{\caption{#1}\vspace{0.5em}#2\if\relax\detokenize{#3}\relax\else\vspace{0.5em}{\footnotesize #3}\fi}

\title{\Large \textbf{Breaking the `Information Silo': Semantic Personas for Cross-Domain Recommendation}}

\author[1]{Jonathan Mayo}
\author[1]{Moshe Unger}
\author[2]{Konstantin Bauman}

\affil[1]{Technology and Information Management Department, Coller School of Management, Tel Aviv University, Tel Aviv 6997801, Israel\\
\texttt{jonathanmayo@mail.tau.ac.il}; \texttt{mosheunger@tauex.tau.ac.il}}
\affil[2]{Management Information Systems Department, Fox School of Business, Temple University, Philadelphia, Pennsylvania 19122\\
\texttt{kbauman@temple.edu}}

\date{\vspace{-1em}}

\begin{document}

\maketitle

\begin{abstract}

Digital platforms increasingly operate as isolated information silos, limiting their ability to construct comprehensive user representations across domains. Cross-domain recommender systems seek to overcome this limitation by transferring knowledge from a source domain to a target domain, yet most existing approaches depend on shared users, shared items, or structurally similar interaction graphs. These assumptions are often unrealistic across independent platforms. We propose SPHERE (Semantic Personas for Heterogeneous cross-domain Recommendation), a design artifact that enables recommendation knowledge transfer across strictly disjoint domains with no shared users or items. Rather than aligning domains through identity or graph structure, SPHERE uses large language models to induce a shared behavioral vocabulary, generate structured semantic personas for users, and retrieve behaviorally similar source-domain communities that form a Community Source Persona. This semantic signal is integrated with collaborative signals through a dual-tower architecture and dynamic fusion gate, allowing SPHERE to augment standard recommender backbones. Empirical evaluation across Amazon Books, Goodreads, and Steam demonstrates consistent improvements over NCF, SVD++, and LightGCN baselines under full-ranking evaluation. The results show that cross-domain transfer effectiveness is not determined solely by semantic proximity between domains; rather, it depends critically on the structural density and native predictive strength of the target domain. The study contributes to information systems research by reframing cross-domain personalization as behavior-based semantic alignment, offering a practical mechanism for overcoming information silos while preserving interpretability and modularity.
\end{abstract}

{Keywords:} recommender systems; cross-domain recommendation; information silos; semantic personas; large language models; personalization; design artifact.

\section{Introduction}\label{sec:Intro}
Recommender Systems (RSes) play a central role in shaping consumer decisions across digital platforms, influencing how users discover products, media, and services across diverse media such as e-commerce, streaming, and travel. In practice, users rarely interact with a single platform in isolation. A person who purchases books online may also stream films, play video games, and browse fashion retailers. These cross-platform behavioral patterns are not independent, random events, but rather manifestations of underlying preference characteristics, such as aesthetic sensibilities, tolerance for complexity, and intrinsic motivations, that transcend any single platform \citep{zang2022survey}. If these macroscopic patterns could be harnessed effectively, RSes could capture comprehensive user representations, thereby substantially improving predictive performance.

Despite this theoretical potential, the development of most RSes remains restricted to independent platforms, yielding an inherently fragmented view of overarching user behavior \citep{cantador2015cross}. This fragmentation, widely termed the 'information silo' effect, a state wherein an information system is incapable of reciprocal operation with inherently related systems \citep{yang2020federated}, constitutes a significant barrier to high-fidelity personalization \citep{shapira2013facebook}. Cross-Domain Recommender Systems (CDRS) aim to address this limitation by transferring knowledge from data-rich source domains to enrich user representations in sparse target domains \citep{khan2017cross}. Two domains are considered complementary when preferences in one provide meaningful predictive signals for the other, even if their respective item catalogs are strictly disjoint \citep{fernandez2012cross}.

However, the efficacy of traditional CDRS architectures is fundamentally constrained by what is increasingly recognized as a 'closed-world assumption' \citep{xu2024rethinking}. Historically, knowledge transfer required explicit entity overlap, such as shared users or deterministic demographic linkages, to establish mathematical anchor points between domains. This strict requirement creates a profound operational problem. In real-world deployments across independent corporate platforms, deterministic overlap is exceptionally rare. While non-generative methodologies have attempted to bypass explicit identifiers by mathematically aligning the topological structures of the source and target domains, these approaches implicitly assume that the two domains share a highly isomorphic geometric structure \citep{zang2022survey}. Consequently, when deployed across highly heterogeneous, strictly disjoint domains (e.g., transferring knowledge from a dense gaming platform to a sparse literary marketplace), purely structural alignments frequently suffer from severe alignment degradation, rendering them operationally unviable \citep{shehmir2025llm4rec}.

To bypass this structural overlap issue, the field has recently witnessed a paradigm shift toward utilizing Large Language Models (LLMs) to bridge disjoint domains via semantic text \citep{shehmir2025llm4rec}. Early adaptations attempted to deploy LLMs as direct, autoregressive ranking engines, prompting the model with a user's source-domain history to predict specific target-domain items \citep{petruzzelli2024instructing}. However, forcing an LLM to act as a sequential ranker introduces severe architectural mismatches. LLMs exhibit inherent computational limitations when scaling to massive item repositories, struggle to extract deep, multi-hop collaborative signals, and frequently succumb to acute hallucination and popularity bias when the source and target domains exhibit large semantic gaps \citep{shehmir2025llm4rec}.

To overcome these limitations, the paradigm must shift from using LLMs as opaque rankers to utilizing them as qualitative, domain-agnostic profiling agents. While recent literature has explored summarizing raw item texts or generating predefined marketing labels to construct user profiles, these methods remain inextricably tied to literal semantic descriptions (e.g., product synopses) \citep{tan2023user}. What is required for true cross-domain knowledge transfer is a deeper behavioral abstraction that captures the universal behavioral drivers underlying consumption across any medium. This fundamental gap motivates our central research question: \textit{how can user preferences be represented in a shared semantic space that supports effective recommendation knowledge transfer across structurally disjoint domains, in which no users or items are shared?} Resolving this requires a fundamental shift from identity-centric or purely structural alignments towards behavior-based representations.

In this paper, we propose the \textbf{Se}mantic \textbf{P}ersonas for \textbf{He}\-tero\-geneous cross-domain \textbf{Re}\-com\-men\-da\-tion (SPHERE) framework, which bridges disjoint domains by first establishing a universal behavioral vocabulary. To ensure that interaction patterns from different platforms can be aligned, SPHERE introduces a data-driven inductive discovery stage. By analyzing representative interaction histories across the paired domains, an LLM extracts a shared taxonomy of behavioral traits. This shared language guarantees that subsequent representations are evaluated along identical behavioral axes. SPHERE then leverages this common taxonomy to generate structured, individual semantic personas for all users. By projecting these trait-aligned personas into a continuous semantic vector space and retrieving similar, but not identical, cross-domain neighbors and aggregating them into a community, the framework enables highly contextualized knowledge transfer which enriches the target user, even when domains share absolutely no structural intersection.

This work contributes to the literature in three principal ways. First, we offer a conceptual reframing of the cross-domain transfer problem, shifting the basis of alignment from identity (shared entities) to behavior (interpretable preference abstractions). By demonstrating that holistic personas can bridge two domains without structural intersection, we directly challenge the pervasive assumption that entity-level overlap is a necessary precondition for effective transfer. Second, we instantiate this reframing as the SPHERE design artifact. Its key elements include the inductive trait discovery procedure, the creation of behavioral personas and the aggregation of a Community Source Persona via semantic neighbor retrieval. The framework functions as a plug-and-play module, which we successfully demonstrate with NCF, SVD++, and LightGCN backbones. Third, our empirical evaluation across Amazon Books, Goodreads, and Steam refines prevailing intuition regarding cross-domain transfer: we demonstrate that more semantically distant pairings (e.g., gaming to literature) do not systematically underperform proximate pairings. Rather, the empirical evidence establishes that transfer efficacy, and the learned fusion gate's reliance on the semantic personas, are governed primarily by target-domain structural density. These results provide concrete design knowledge for building interpretable and robust recommender systems across isolated platforms.

\section{Related Work}\label{sec:Related}

\subsection{Cross-Domain Recommendation and the Overlap Dilemma}\label{sec:overlap_dilemma}
Recommender systems frequently operate within isolated environments, 'Information Silos', encountering severe data sparsity challenges where insufficient interaction history impedes accurate user modeling. To mitigate this, CDRS frameworks transfer collaborative signals from data-rich source domains to sparse target domains. The architectural progression of CDRS began with explicit bridging mechanisms, such as EMCDR \citep{man2017cross}, which pioneered embedding mapping functions. Subsequent models, including GA-DTCDR \citep{zhu2020graphical}, introduced graph-based mechanisms to enhance relational representations, while methodologies such as SEAGULL \citep{zhao2023cross} and CVPM \citep{zhao2025cross} attempted to relax strict overlap through progressive structural alignment. In extreme sparsity settings, frameworks like KR-CDR \citep{huang2024knowledge} incorporated auxiliary knowledge graphs, and ARISEN \citep{zhao2023sequential} aligned time-series sequences. However, these foundational frameworks relied heavily on deterministic entity linkages to establish cross-domain anchor points. Recognizing that explicit overlap is operationally infeasible across independent corporate silos, the field encountered a critical 'overlap dilemma' \citep{liu2025bridge}: relying on overlapping users severely compromises the practicability of recommendation frameworks in the real world.

To bypass this dilemma, subsequent methodologies attempted to align disjoint domains structurally, treating latent spaces as distinct probability distributions. Techniques utilizing multimodal domain adaptation \citep{shyam2025cross} and zero-shot popularity dynamics \citep{wang2024pre} sought to circumvent explicit ID mapping. Parallel efforts in unsupervised latent alignment \citep{wang2024making} and Multi-view Graph Contrastive Learning \citep{liu2024extracting} forced source and target distributions to match mathematically. While these statistical approaches eliminate the need for shared identifiers, they rely heavily on the assumption that the geometric topologies of the two domains are fundamentally isomorphic. Consequently, when deployed across highly heterogeneous domains with disparate interaction densities, these structural alignments frequently degrade due to irreconcilable geometric mismatch between the source and target manifolds \citep{shehmir2025llm4rec}.

To overcome geometric warping, recent models pivoted toward utilizing continuous semantic representations derived from item texts. Frameworks such as FFMSR \citep{lu2025federated} construct global semantic prototypes through the static clustering of item descriptions, while PFCR \citep{guo2024prompt} utilizes vector quantization, and FedCRF \citep{guo2026fedcrf} dynamically re-estimates semantic cluster centers. More recently, sophisticated sequence-level architectures like X-Cross \citep{hadad2025x} have integrated language models to bridge cross-domain sequences dynamically. While these semantic-driven models successfully bypass structural isomorphism, they remain constrained by item-level granularity. Sequence-level bridging struggles across massive semantic gaps where the raw items share zero conceptual overlap (e.g., aligning interactions between a computer mousepad and a romantic comedy film). This fundamental limitation necessitates an abstraction from item-level semantics to user-level behavior.

\subsection{Explicit Generative User Modeling and Behavioral Personas}\label{sec:llms_personas}
The transition toward user-level abstraction aligns with a broader shift in the literature towards 'Explicit Generative User Modeling' \citep{tan2023user}. Historically, efforts to bridge domains via user profiles relied on extracting latent traits using demographic data or tag-based matrices \citep{sahu2019user}. While these foundational methods successfully established the precedent of utilizing user profiles as cross-domain bridges to reduce sparsity, they were fundamentally constrained by pre-defined vocabularies and structural rigidity, rendering them incapable of zero-shot generalization across highly disparate domains. 

The integration of LLMs naturally resolved this rigidity, but initial implementations deployed them as direct, autoregressive ranking engines \citep{petruzzelli2024instructing}. As established by \cite{xin2025llmcdsr}, forcing an LLM to predict specific items introduces severe operational mismatches, as LLMs struggle to handle extensive item repositories and fail to extract deep collaborative signals. Furthermore, direct-prompting approaches suffer from acute hallucination when source and target domains exhibit large semantic gaps \citep{liu2025uncovering}.

Recognizing the limitations of direct ranking, state-of-the-art frameworks pivoted toward utilizing LLMs as continuous feature extractors and profilers. Models like User-LLM \citep{ning2025user} address the context-window limitation by compressing user histories into continuous embeddings (soft prompts) for efficient contextualization, though at the complete loss of human interpretability. Conversely, textual approaches have sought to restore this interpretability. Frameworks such as SAGCN \citep{liu2025understanding} emphasize the principle of 'understanding before recommendation' by leveraging LLMs to extract fine-grained semantic aspects from user reviews, a concept expanded upon by LIPA \citep{azam2026reviews} and ONE \citep{tang2026one}, which summarize item texts into domain-agnostic profiles. Concurrently, models targeting e-commerce have generated explicit, human-readable customer personas \citep{shi2025you}. While these approaches successfully decouple the LLM from the ranking task, they remain constrained by semantic literalism. They rely on predefined marketing labels (e.g., 'Bargain Hunter') or item-level aspect summaries that fail to transfer accurately across heterogeneous mediums. 

What is required for true heterogeneous transfer is a domain-agnostic behavioral abstraction. Recent empirical studies at the intersection of Human-Computer Interaction and psychology have demonstrated that LLMs possess the zero-shot reasoning capabilities necessary to accurately deduce stable, validated psychological traits (e.g., cognitive styles and locus of control) strictly from sparse, unstructured digital footprints \citep{schuller2024generating}. 

Building upon the principle of semantic extraction prior to recommendation \citep{liu2025understanding}, SPHERE extends this line of inquiry by introducing an inductive, data-driven discovery phase. Crucially, rather than extracting granular, item-specific review aspects, the LLM analyzes the cross-domain corpus to discern a universal taxonomy of behavioral traits, thereby establishing a shared semantic vocabulary. This substantially exceeds the rigid profile-extraction capabilities of early pre-LLM frameworks \citep{sahu2019user}. SPHERE then generates structured semantic personas grounded exclusively in these abstract traits. To make this natural language mathematically actionable, SPHERE utilizes modern dense sentence embedding models \citep{neelakantan2022text}, pre-trained using contrastive learning objectives, to project these behavioral personas into a continuous vector space ($\mathbb{R}^d$). This ensures the user's nuanced preferences and characteristics are mathematically preserved, allowing standard collaborative filtering architectures to bypass ID matching entirely and dynamically identify cross-domain behavioral neighbors.

\section{Methodology}\label{sec:Method}

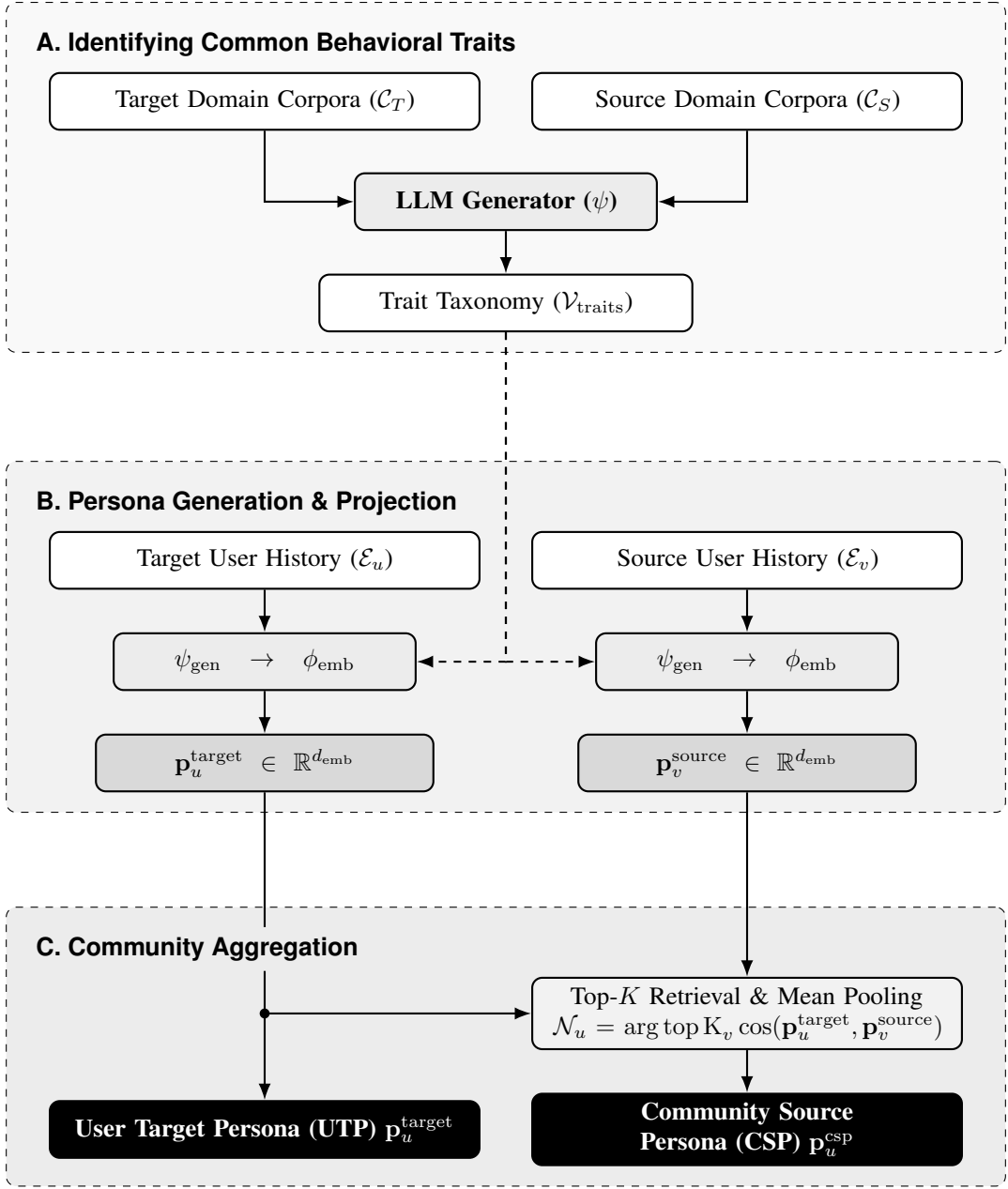
\begin{figure}[htbp]
\centering
\begin{tikzpicture}[
    >=Latex,
    box/.style={rectangle, draw=black, thick, fill=white,
                text width=5.8cm, align=center, rounded corners,
                minimum height=0.8cm, font=\small},
    proc/.style={rectangle, draw=black, thick, fill=gray!15,
                 text width=4cm, align=center, rounded corners,
                 minimum height=0.8cm, font=\small\bfseries},
    wide/.style={rectangle, draw=black, thick, fill=white,
                 text width=8.5cm, align=center, rounded corners,
                 minimum height=0.8cm, font=\small},
    vec/.style={rectangle, draw=black, thick, fill=gray!30,
                text width=4.5cm, align=center, rounded corners,
                minimum height=0.8cm, font=\small\bfseries},
    csp/.style={rectangle, draw=black, thick, fill=black,
                text=white, text width=5.8cm, align=center,
                rounded corners, minimum height=0.8cm,
                font=\small\bfseries},
    arrow/.style={draw, thick, ->, >=Latex},
    dashed_arrow/.style={draw, thick, dashed, ->, >=Latex},
    panel_label/.style={font=\sffamily\bfseries\small, anchor=north west, 
                        rounded corners=2pt, inner sep=4pt},
    panel_bg/.style={draw=black, dashed, rounded corners, inner xsep=0pt, inner ysep=8pt}
]

\node[box] (target_corp) at (-3.4, 0) {Target Domain Corpora ($\mathcal{C}_T$)};
\node[box] (source_corp) at (3.4, 0) {Source Domain Corpora ($\mathcal{C}_S$)};

\path (target_corp.east) -- (source_corp.west) coordinate[midway] (mid_corp);

\node[proc, below=1cm of mid_corp] (llm_disc) {LLM Generator ($\psi$)};
\node[wide, text width=5cm, below=0.6cm of llm_disc] (taxonomy) {Trait Taxonomy ($\mathcal{V}_{\mathrm{traits}}$)};

\draw[arrow] (target_corp.south) |- (llm_disc.west);
\draw[arrow] (source_corp.south) |- (llm_disc.east);
\draw[arrow] (llm_disc) -- (taxonomy);

\node[box, below=2.8cm of taxonomy, xshift=-3.4cm] (target_hist) {Target User History ($\mathcal{E}_u$)};
\node[box, below=2.8cm of taxonomy, xshift=3.4cm] (source_hist) {Source User History ($\mathcal{E}_v$)};

\node[proc, below=0.6cm of target_hist] (pipe_t) {$\psi_{\mathrm{gen}} \;\rightarrow\; \phi_{\mathrm{emb}}$};
\node[proc, below=0.6cm of source_hist] (pipe_s) {$\psi_{\mathrm{gen}} \;\rightarrow\; \phi_{\mathrm{emb}}$};

\node[vec, below=0.6cm of pipe_t] (vec_t) {$\mathbf{p}_u^{\mathrm{target}} \in \mathbb{R}^{d_{\mathrm{emb}}}$};
\node[vec, below=0.6cm of pipe_s] (vec_s) {$\mathbf{p}_v^{\mathrm{source}} \in \mathbb{R}^{d_{\mathrm{emb}}}$};

\draw[arrow] (target_hist) -- (pipe_t);
\draw[arrow] (source_hist) -- (pipe_s);
\draw[arrow] (pipe_t) -- (vec_t);
\draw[arrow] (pipe_s) -- (vec_s);

\draw[dashed_arrow] (taxonomy.south) |- (pipe_t.east);
\draw[dashed_arrow] (taxonomy.south) |- (pipe_s.west);


\node[wide, fill=gray!10, text width=5.8cm, below=2.6cm of vec_s] (knn) {Top-$K$ Retrieval \& Mean Pooling\\ \normalfont $\mathcal{N}_u = \operatorname*{arg\,top\,K}_{v} \cos(\mathbf{p}_u^{\mathrm{target}}, \mathbf{p}_v^{\mathrm{source}})$};

\node[csp, below=0.6cm of knn] (csp_out) {Community Source Persona (CSP) $\mathbf{p}_u^{\mathrm{csp}}$};

\node[csp, text width=5.8cm] (utp_out) at (vec_t |- csp_out) {User Target Persona (UTP) $\mathbf{p}_u^{\mathrm{target}}$};

\draw[arrow] (vec_t.south) -- (utp_out.north);
\draw[arrow] (vec_s.south) -- (knn.north);
\draw[arrow] (knn.south) -- (csp_out.north);

\coordinate (query_split) at (vec_t.south |- knn.west);
\fill (query_split) circle (2pt); 
\draw[arrow] (query_split) -- (knn.west);

\path (target_corp.west) -- ++(-0.6,0) coordinate (Left);
\path (source_corp.east) -- ++(0.6,0) coordinate (Right);

\path (target_corp.north) ++(0, 0.7) coordinate (A_top);
\path (target_hist.north) ++(0, 0.7) coordinate (B_top);
\path (knn.north) ++(0, 0.7) coordinate (C_top); 

\begin{scope}[on background layer]
    \node[panel_bg, fill=gray!5, fit=(Left |- A_top) (Right |- taxonomy.south)] (panelA) {};
    
    \node[panel_bg, fill=gray!10, fit=(Left |- B_top) (Right |- vec_t.south)] (panelB) {};
    
    \node[panel_bg, fill=gray!15, fit=(Left |- C_top) (Right |- csp_out.south)] (panelC) {};
\end{scope}

\node[panel_label, fill=gray!5] at ([xshift=8pt, yshift=-8pt]panelA.north west) {A. Identifying Common Behavioral Traits};
\node[panel_label, fill=gray!10] at ([xshift=8pt, yshift=-8pt]panelB.north west) {B. Persona Generation \& Projection};
\node[panel_label, fill=gray!15] at ([xshift=8pt, yshift=-8pt]panelC.north west) {C. Community Aggregation};

\end{tikzpicture}
\caption{The offline semantic persona pipeline. The framework inductively discovers a shared behavioral vocabulary from cross-domain corpora~(A), generates and embeds structured personas for all users conditioned on this taxonomy~(B), and utilizes the target persona to retrieve and aggregate the $K$ nearest source-domain neighbors per user into the Community Source Persona (CSP). Both the CSP and the isolated User Target Persona (UTP) are extracted for downstream evaluation within the dual-tower architecture (Figure~\ref{fig:sphere_architecture})~(C).}
\label{fig:sphere_overview}
\end{figure}

In this section, we introduce the \textbf{Se}mantic \textbf{P}ersonas for \textbf{He}\-tero\-geneous cross-domain \textbf{Re}\-com\-men\-da\-tion (SPHERE) framework, illustrated in Figure \ref{fig:sphere_overview}. SPHERE addresses a key limitation of existing cross-domain recommender systems: their reliance on structural overlap, such as shared users, shared items, or closely related item representations. Rather than aligning domains through common entities, SPHERE aligns them through LLM-generated semantic personas that summarize users’ domain-agnostic behavioral tendencies.

SPHERE is instantiated as a recommender-system design artifact and implemented as a semantic augmentation module for existing recommender backbones, including NCF, SVD++, and LightGCN. Its dual-tower architecture combines target-domain collaborative signals with source-domain behavioral personas. The framework first induces a shared taxonomy of behavioral traits from representative interaction histories, uses this taxonomy to generate structured personas for individual users, and retrieves behaviorally similar cross-domain neighbors to construct a \emph{Community Source Persona} (CSP). This persona-based representation enriches target-user representations and enables transfer across domains with no shared users, items, or interaction structure.

The remainder of this section is organized as follows. Section \ref{sec:preliminaries} formalizes the cross-domain setting and introduces the semantic bridge formulation. Sections \ref{sec:trait_discovery}--\ref{sec:csp} describe the three core stages of persona-based transfer: identifying common behavioral traits, generating semantic personas, and constructing the CSP for knowledge transfer. Section \ref{sec:architecture} then presents the SPHERE architecture. Specifically, Sections \ref{sec:collab_encoding} and \ref{sec:semantic_encoding} describe the collaborative and semantic encoding towers, Section \ref{sec:fusion} introduces the dynamic late-fusion and scoring mechanism, and Section \ref{sec:training} details the model training procedure and optimization objective.

\subsection{Preliminaries and Semantic Bridge Formulation}\label{sec:preliminaries}

We establish a standardized mathematical notation for the SPHERE framework, consolidated in Table \ref{tab:notation}. Let $\mathcal{U}_T$ and $\mathcal{U}_S$ denote the strictly disjoint user sets in the target domain $\mathcal{D}_T$ and the source domain $\mathcal{D}_S$, respectively ($\mathcal{U}_T \cap \mathcal{U}_S = \emptyset$). Let $\mathcal{I}_T$ and $\mathcal{I}_S$ represent the respective item sets. While item overlap may exist across domains in practice, our framework operates without assuming or utilizing any structural intersection. For a given user $u \in \mathcal{U}_T$ in the target domain, let $\mathcal{H}_u = \{i_1, i_2, \dots, i_n\}$ represent the set of historical interactions, where $i_k \in \mathcal{I}_T$.

\begin{table}[htbp]
\TABLE
{Notation Table\label{tab:notation}}
{
\begin{tabular}{p{0.15\textwidth} p{0.8\textwidth}}
\hline\up 
\textbf{Notation} & \textbf{Description} \\ \hline\up 
$\mathcal{D}_T, \mathcal{D}_S$ & The target domain and source domain, respectively. \\
$\mathcal{U}_T, \mathcal{U}_S$ & Disjoint sets of users in the target domain ($\mathcal{D}_T$) and source domain ($\mathcal{D}_S$). \\
$\mathcal{I}_T, \mathcal{I}_S$ & Sets of items within the target and source domains. \\
$\mathcal{H}_u$ & Set of historical interactions for a user $u$ within their respective domain. \\
$\mathcal{E}_u$ & Metadata-enriched interaction history for user $u$ (titles, genres, summaries). \\
$M$ & The number of dynamic, inductively discovered bridging traits. \\
$\mathcal{V}_{\mathrm{traits}}$ & The universal trait taxonomy inductively discovered by the LLM. \\
$\mathbf{p}_u^{\mathrm{target}}$ & The generated individual textual persona embedding for target user $u$ ($\mathbf{p}_u^{\mathrm{target}} \in \mathbb{R}^{d_{\mathrm{emb}}}$). \\
$\mathbf{p}_v^{\mathrm{source}}$ & The generated individual textual persona embedding for source user $v$ ($\mathbf{p}_v^{\mathrm{source}} \in \mathbb{R}^{d_{\mathrm{emb}}}$). \\
$\mathcal{N}_u$ & The subset of $K$ semantic neighbors in the source domain for a target user $u$. \\
$\mathbf{p}_u^{\mathrm{csp}}$ & The aggregated Community Source Persona embedding. \\
$\mathbf{z}_u, \mathbf{z}_i$ & Latent representations of user $u$ and item $i$ from the collaborative backbone. \\
$\mathbf{e}_i^{\mathrm{sem}}$ & The dedicated semantic representation of candidate item $i$ ($\mathbf{e}_i^{\mathrm{sem}} \in \mathbb{R}^d$). \\
$\mathbf{h}_{u,i}^{\mathrm{col}}$ & The joint non-linear structural interaction vector. \\
$\mathbf{h}_{u,i}^{\mathrm{sem}}$ & The joint non-linear semantic behavioral interaction vector. \\
$\alpha_{u,i}$ & The instance-aware scalar weight computed by the late-fusion gate. \\
$\phi_{\mathrm{emb}}$ & The dense embedding model utilized to project textual personas into the continuous vector space. \\
$\psi, \psi_{\mathrm{gen}}$ & The LLM utilized as a stochastic generative process ($\psi$) and structured text generator ($\psi_{\mathrm{gen}}$). \\
$\rho$ & The prompt formulation function constraining the generation process. \\
$\tau$ & The learnable temperature parameter for gradient stability. \\
$K$ & The semantic community size for aggregating the source persona. \\
$N$ & The number of unobserved negative items sampled during contrastive optimization. \\
$d_{\mathrm{emb}}$ & Dimensionality of the dense semantic embedding space output by $\phi_{\mathrm{emb}}$. \\
$d, d'$ & Dimensionalities of the latent representations and structural/semantic interaction vectors. \down\\ \hline
\end{tabular}
}
{}
\end{table}

To construct a domain-agnostic behavioral persona that can enrich these target-domain representations, SPHERE proceeds in three stages, detailed in Sections~\ref{sec:trait_discovery}--\ref{sec:csp}: an LLM first induces a shared taxonomy of behavioral traits from representative cross-domain interaction data; individual users in both domains are then profiled along these trait axes and projected into a continuous embedding space; finally, for each target user, the $K$ most behaviorally similar source-domain neighbors are retrieved and aggregated into a single Community Source Persona, denoted as $\mathbf{p}_u^{\mathrm{csp}}$. 

Given the user history $\mathcal{H}_u$ and the transferred persona signal $\mathbf{p}_u^{\mathrm{csp}}$, the recommendation task is formalized as Top-$N$ item ranking. We learn a parameterized scoring function $f_\Theta(u, i \mid \mathbf{p}_u^{\mathrm{csp}})$ that computes a relative affinity score between user $u$ and an unobserved candidate item $i \in \mathcal{I}_T \setminus \mathcal{H}_u$, optimized to rank true interactions above negative candidates by leveraging the semantic behavioral persona derived from $\mathcal{D}_S$, thereby improving ranking accuracy in the target domain without relying on direct user or item overlap.

\subsubsection{Identifying Common Behavioral Traits}\label{sec:trait_discovery}\mbox{}\newline
To establish a universal semantic vocabulary capable of bridging structurally disjoint domains, we first perform a data-driven trait discovery phase. Rather than relying on predefined, domain-specific feature taxonomies, we leverage a Large Language Model (LLM) to inductively identify shared behavioral dimensions from interaction data across the source and target domains \citep{ebrat2026end}. These common traits serve as a shared representational vocabulary, enabling the LLM to generate persona texts for users in both domains along the same behavioral dimensions. As a result, source- and target-domain users can be projected into a comparable semantic space despite the absence of shared users, items, or interaction structures.

For each user in the target and source domains, we extract the textual metadata associated with every item in their interaction history: specifically, the item's title, genre tags, and descriptive summary. These metadata-enriched interaction histories are aggregated to form domain-level corpora, denoted $\mathcal{C}_T$ and $\mathcal{C}_S$. For example, a single entry in a literary domain corpus might comprise: \textit{`Book Name'}, [Genres List], \textit{`Synopsis'}.

We model the LLM as a stochastic generative process $\psi(\cdot)$, instructed to identify latent behavioral patterns that generalize across domain-specific item metadata. By jointly processing the aggregated interaction patterns within both corpora, the LLM derives a taxonomy of $M$ bridging traits:
\begin{equation}\label{eq:traits}
\mathcal{V}_{\mathrm{traits}} = \{t_1, t_2, \dots, t_M\} = \psi(\operatorname{concat}(\mathcal{C}_T, \mathcal{C}_S))
\end{equation}
where $\operatorname{concat}(\cdot)$ denotes the prompt-based textual aggregation of the respective domain corpora, and each $t_m$ represents a qualitative, domain-agnostic behavioral dimension.

Crucially, the framework does not impose a rigid, predetermined limit on the number of traits. Instead, $M$ remains dynamic, allowing the LLM to expand or contract the trait taxonomy based on the behavioral complexity of the specific source--target domain pair. This design choice is central to SPHERE because the semantic bridge required for cross-domain transfer is not universal; it depends on the behavioral relationship between the domains being connected. Closely related domains, such as Goodreads and Amazon Books, can often be aligned through a relatively compact set of literary preference dimensions, such as cognitive intensity, realism versus fantasy, moral ambiguity, and information-seeking versus escapism. In contrast, more heterogeneous pairings, such as Steam and Amazon Books, require broader behavioral abstractions, including agency, stimulation seeking, challenge orientation, progression motivation, and tolerance for conflict.

SPHERE therefore induces the trait taxonomy from the paired corpora themselves rather than imposing a predefined psychological, marketing, or domain-specific taxonomy. This makes the taxonomy both \textit{shared} and \textit{adaptive}. It is \textit{shared} because the same induced traits are subsequently used to generate personas for users in both the source and target domains, ensuring that users from both domains are represented along comparable behavioral dimensions. It is \textit{adaptive} because the content and granularity of the taxonomy are determined by the behavioral signals that are salient for the specific domain pair.

Table~\ref{tab:prompts} illustrates this design principle by reporting the trait taxonomies induced for each domain configuration. The table shows that literary-domain pairings converge on traits that capture reading-oriented preferences, whereas cross-media pairings involving Steam require more general behavioral dimensions that can connect gaming behavior with literary or product-consumption behavior. These induced traits provide an interpretable behavior-level representation that can bridge domains without requiring shared users, shared items, or structurally similar interaction graphs.

\begin{table*}[htbp]
\TABLE
{Inductively Discovered Behavioral Trait Taxonomies for Semantic Persona Generation\label{tab:prompts}}
{
\begin{tabular}{p{0.2\textwidth} p{0.76\textwidth}}
\hline\up 
\textbf{Context} & \textbf{Behavioral Trait Taxonomy} \\ \hline\up 
\raggedright\textbf{Cross-Domain Bridge} \newline (Steam $\leftrightarrow$ Goodreads/Amazon Books) & 
\raggedright
\textbf{Stimulation:} Need for novelty and high-sensation vs. comfort with familiar routines. \newline
\textbf{Locus:} Desire for internal agency and control vs. willingness to be guided by external forces. \newline
\textbf{Engagement:} Preference for active, systemic problem-solving vs. passive, atmospheric absorption. \newline
\textbf{Investment:} Tolerance for long-term, high-friction commitments vs. desire for quick, low-barrier gratification. \newline
\textbf{Conflict:} Attraction to tension, risk, and antagonism vs. preference for harmony and resolution. \newline
\textbf{Narrative:} Motivation driven by structural mechanics and systems vs. character-driven storytelling. \newline
\textbf{Agency:} Need to express individuality and creativity vs. following optimized, predefined pathways. \tabularnewline \hline
\raggedright\textbf{Literary Domains} \newline (Goodreads/Amazon Books in-domain and Goodreads $\leftrightarrow$ Amazon Books) & 
\raggedright
\textbf{Information Intent:} Escapism vs. knowledge acquisition. \newline
\textbf{Cognitive Intensity:} Preference for complex, demanding structures vs. light processing. \newline
\textbf{Ontological Anchor:} Grounded realism vs. speculative or fantasy abstraction. \newline
\textbf{Moral Spectrum:} Comfort with ambiguity and darkness vs. clear hero/villain binaries. \newline
\textbf{Cultural Scope:} Focus on localized, familiar settings vs. expansive, alien, or global scales. \tabularnewline \hline
\raggedright\textbf{Gaming Domain} \newline (Steam in-domain) & 
\raggedright
\textbf{Primary Driver:} Competition and dominance vs. exploration and creation. \newline
\textbf{Social Friction:} Preference for cooperative, solitary, or antagonistic interactions. \newline
\textbf{Pacing:} Tolerance for slow, methodical execution vs. rapid, twitch-reflex demands. \newline
\textbf{Progression:} Motivation via extrinsic rewards (stats, loot) vs. intrinsic mastery. \newline
\textbf{Thematic Intensity:} Preference for high-stakes, stressful atmospheres vs. relaxing or cozy environments. \newline
\textbf{Production Ethos:} Attraction to polished, mainstream structures vs. experimental or indie designs. \down\tabularnewline \hline
\end{tabular}
}
{}
\end{table*}

\subsection{Persona Generation \& Projection}\label{sec:persona_generation}\mbox{}

Having established the shared trait taxonomy $\mathcal{V}_{\mathrm{traits}}$, SPHERE next replaces purely ID-based user representations with non-identifying semantic personas. Rather than representing a user only through a platform-specific identifier or latent embedding tied to a single interaction matrix, we summarize the user's interaction history as a structured textual profile expressed along the shared behavioral traits. This persona does not encode the user's identity; instead, it captures domain-agnostic behavioral tendencies that can be compared across source and target domains. We then project these textual personas into dense vector representations, placing users from both domains in a common semantic space.

Specifically, to generate these personas, we first compile a natural-language summary of the user's interaction history. Rather than relying on raw item identifiers, we enrich each interacted item with its descriptive metadata---specifically its title, primary genres, and a brief text synopsis. This enriched historical context is then injected into a standardized prompt template, as shown in Figure~\ref{fig:llm_prompt}.

Crucially, instead of allowing the LLM to generate an unconstrained, open-ended profile, the prompt explicitly restricts the model to evaluate the user strictly along the dimensions of the inductively discovered trait taxonomy ($\mathcal{V}_{\mathrm{traits}}$) and enforces a strict media-agnostic output constraint.

\begin{figure}[htbp]
\centering
\fbox{%
    \parbox{\dimexpr\textwidth-2\fboxsep-2\fboxrule\relax}{%
        \small
        \textbf{System Prompt:} You are a behavioral-science pattern-matching engine. Task: Convert user history into a Media-Agnostic psychological profile.\\
        
        \textbf{[RULES]} \\
        1. Begin output EXACTLY with: "The user's behavioral profile indicates" \\
        2. OUTPUT ONLY the profile. No intro, no conversational filler. \\
        3. ABSOLUTELY NO SPECIFIC TITLES, PROPER NOUNS, OR NAMED WORKS. \\
        4. NO SPECIFIC MEDIA TERMS (e.g., no "reader", "player", "book", "game"). \\
        5. EXACTLY 2 SENTENCES. Continuous text.\\
        
        \textbf{[CURRENT TASK]} \\
        Analyze cross-domain behavior using the following standardized clinical framework: \\
        - Stimulation: Need for novelty and high-sensation vs.\ comfort with familiar routines. \\
        - Locus: Desire for internal agency/control vs.\ willingness to be guided by external forces. \\
        - Engagement: Preference for active, systemic problem-solving vs.\ passive, atmospheric absorption. \\
        - Investment: Tolerance for long-term, high-friction commitments vs.\ desire for quick, low-barrier gratification. \\
        - Conflict: Attraction to tension, risk, and antagonism vs.\ preference for harmony and resolution. \\
        - Narrative: Motivation driven by structural mechanics/systems vs.\ character-driven storytelling. \\
        - Agency: Need to express individuality/creativity vs.\ following optimized, predefined pathways.\\
        
        \textbf{[USER HISTORY]} \\
        \textit{`East of Eden'}, [Fiction, Family Life, Multi-generational], \textit{`Set in the rich farmland of California's Salinas Valley, this sprawling and often brutal novel follows the intertwined destinies of two families whose generations helplessly reenact the fall of Adam and Eve and the poisonous rivalry of Cain and Abel.'}\\
        \textit{`Book 2'}, [Genre], \textit{`Synopsis'} \\
        \vdots \\
        \textit{[Full interaction history provided]}
    }%
}
\caption{The standardized LLM prompt template utilized to generate media-agnostic behavioral personas.}
\label{fig:llm_prompt}
\end{figure}

By strictly conditioning the generation process on these predefined rules and the $\mathcal{V}_{\mathrm{traits}}$ taxonomy, this prompt acts as a rigorous semantic constraint. It guarantees that the resulting textual profile for a target user and the profile for a source user are evaluated along the exact same behavioral axes without relying on domain-specific terminology. This structural alignment ensures that when these texts are subsequently projected into the dense vector space, they reside in a comparable semantic subspace, rendering the representations entirely independent of the specific, disjoint item IDs.

To execute this, we employ the LLM---utilizing the same underlying model from the discovery phase---operating here as a structured text generation function $\psi_{\mathrm{gen}}(\cdot)$. The resulting qualitative behavioral profile is then projected into a $d_{\mathrm{emb}}$-dimensional continuous vector space using a pre-trained language model $\phi_{\mathrm{emb}}(\cdot)$, such as BERT. 

Let $\rho(\mathcal{E}_u, \mathcal{V}_{\mathrm{traits}})$ denote the formatted prompt combining the user's enriched history and the taxonomy constraint. Formally, the individual target persona embedding is computed as:
\begin{equation}\label{eq:target_persona}
\mathbf{p}_u^{\mathrm{target}} = \phi_{\mathrm{emb}}(\psi_{\mathrm{gen}}(\rho(\mathcal{E}_u, \mathcal{V}_{\mathrm{traits}})))
\end{equation}

Similarly, for every user $v \in \mathcal{U}_S$ in the source domain with an enriched interaction history $\mathcal{E}_v$, we independently compute the individual source persona embedding:
\begin{equation}\label{eq:source_persona}
\mathbf{p}_v^{\mathrm{source}} = \phi_{\mathrm{emb}}(\psi_{\mathrm{gen}}(\rho(\mathcal{E}_v, \mathcal{V}_{\mathrm{traits}})))
\end{equation}
The output of this phase is a set of persona embeddings for users in both domains. Specifically, each target-domain user $u \in \mathcal{U}_T$ and each source-domain user $v \in \mathcal{U}_S$ is represented by a structured persona generated along the same trait taxonomy $\mathcal{V}_{\mathrm{traits}}$ and then projected into the shared semantic space. Thus, the source and target personas are not arbitrary textual summaries, but trait-conditioned representations constructed using the same behavioral vocabulary. This shared structure enables meaningful comparison between users across domains while decoupling user preference representation from domain-specific item identifiers.

\subsubsection{Community Aggregation – Creating the Community Source Persona (CSP)}\label{sec:csp}\mbox{}\newline
The core of SPHERE's cross-domain transfer strategy is to construct, for each target user, an aggregated source-domain behavioral persona. Because the source and target domains do not share user identities, SPHERE does not attempt to find the same user across domains. Instead, it uses the target user's persona as a semantic query to retrieve source-domain users who express similar behavioral traits. For each target user $u$, we identify a set of $K$ semantic neighbors $\mathcal{N}_u$ by maximizing the cosine similarity between their respective individual persona embeddings:
\begin{equation}\label{eq:neighbors}
\mathcal{N}_u = \operatorname*{arg\,top\,K}_{v \in \mathcal{U}_S} \cos(\mathbf{p}_u^{\mathrm{target}}, \mathbf{p}_v^{\mathrm{source}})
\end{equation}

We define an aggregation function $\Phi$ that synthesizes the representations of the $K$ retrieved neighbors into a single dense vector. Specifically, we employ an unweighted mean pooling across the neighborhood. The Community Source Persona is thus formally computed as:
\begin{equation}\label{eq:csp}
\mathbf{p}_u^{\mathrm{csp}} = \frac{1}{K} \sum_{v \in \mathcal{N}_u} \mathbf{p}_v^{\mathrm{source}}
\end{equation}

We deliberately utilize an unweighted mean rather than a similarity-weighted average due to the geometric properties of high-dimensional embedding spaces. In highly dense vector spaces, cosine similarity distributions exhibit extreme top-edge concentration. Consequently, the variance in similarity scores among a tightly clustered set of top-$K$ neighbors is mathematically marginal. In this regime, an unweighted mean provides a highly stable, computationally efficient representation of the community's behavioral center without introducing the noise associated with scaling by functionally identical similarity weights \citep{ethayarajh2019contextual}.

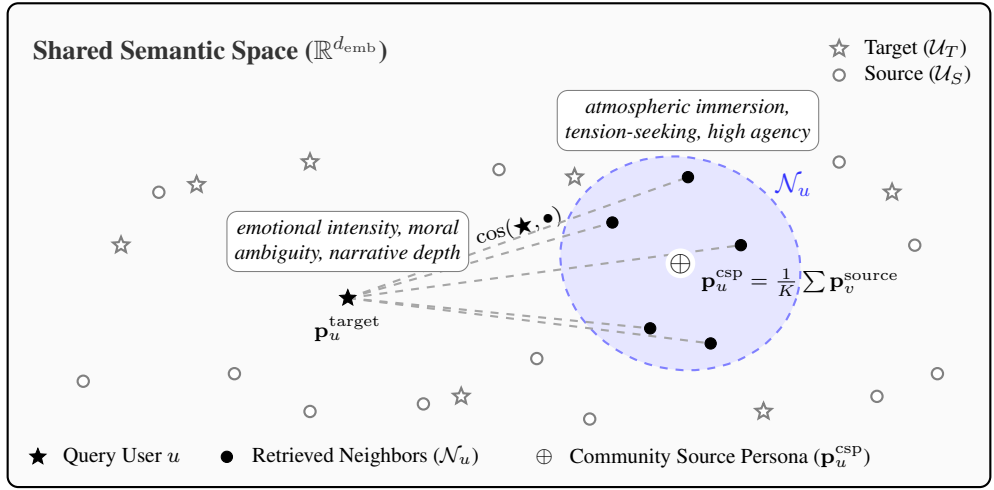
\begin{figure}[htbp]
\centering
\begin{tikzpicture}[
    >=Latex,
    target/.style={star, star points=5, star point ratio=2.25, draw=black, fill=black, inner sep=1.0pt},
    unselected_target/.style={star, star points=5, star point ratio=2.25, draw=gray, fill=white, thick, inner sep=1.0pt},
    selected/.style={circle, draw=black, fill=black, inner sep=1.5pt},
    unselected/.style={circle, draw=gray, fill=white, thick, inner sep=1.5pt},
    callout/.style={rectangle, draw=gray, fill=white, rounded corners, font=\scriptsize, align=center, inner sep=3pt, outer sep=2pt},
    font=\sffamily
]

\draw[thick, rounded corners, fill=gray!4] (-6.5, -3.2) rectangle (6.5, 3.2);
\node[anchor=north west, font=\bfseries\small, text=black!80] at (-6.3, 2.9) {Shared Semantic Space ($\mathbb{R}^{d_{\mathrm{emb}}}$)};

\node[unselected_target] at (4.5, 2.6) {}; 
\node[anchor=west, font=\scriptsize] at (4.7, 2.6) {Target ($\mathcal{U}_T$)};

\node[unselected] at (4.5, 2.25) {}; 
\node[anchor=west, font=\scriptsize] at (4.7, 2.25) {Source ($\mathcal{U}_S$)};

\begin{scope}[yshift=-0.5cm]

    \coordinate (T) at (-2.0, -0.2);
    
    \coordinate (N1) at (1.5, 0.8);
    \coordinate (N2) at (2.5, 1.4);
    \coordinate (N3) at (2.0, -0.6);
    \coordinate (N4) at (3.2, 0.5);
    \coordinate (N5) at (2.8, -0.8);
    
    \coordinate (C) at (2.4, 0.26);
    
    \draw[draw=blue!50, fill=blue!10, thick, dashed, rotate around={-15:(C)}] (C) ellipse (1.6cm and 1.4cm);
    \node[font=\small\bfseries, text=blue!80] at (3.9, 1.3) {$\mathcal{N}_u$};
    
    \node[unselected] at (-4.5, 1.2) {};
    \node[unselected] at (-3.5, -1.2) {};
    \node[unselected] at (-1.0, -1.6) {};
    \node[unselected] at (0.0, 1.5) {};
    \node[unselected] at (0.5, -1.0) {};
    \node[unselected] at (4.5, 1.6) {};
    \node[unselected] at (5.0, -1.5) {};
    \node[unselected] at (5.5, 0.5) {};
    \node[unselected] at (-5.5, -1.3) {};
    \node[unselected] at (-2.5, -1.7) {};
    \node[unselected] at (5.8, -1.2) {};
    \node[unselected] at (1.2, -1.8) {};
    
    \node[unselected_target] at (-5.0, 0.5) {};
    \node[unselected_target] at (-2.5, 1.6) {};
    \node[unselected_target] at (-0.5, -1.5) {};
    \node[unselected_target] at (3.5, -1.7) {};
    \node[unselected_target] at (1.0, 1.4) {};
    \node[unselected_target] at (-4.0, 1.3) {};
    \node[unselected_target] at (5.2, 1.2) {};
    
    \draw[dashed, draw=gray!70, thick] (T) -- (N1) node[pos=0.67, sloped, above, font=\scriptsize, text=black] {$\cos(\bigstar, \bullet)$};
    \draw[dashed, draw=gray!70, thick] (T) -- (N2);
    \draw[dashed, draw=gray!70, thick] (T) -- (N3);
    \draw[dashed, draw=gray!70, thick] (T) -- (N4);
    \draw[dashed, draw=gray!70, thick] (T) -- (N5);
    
    \node[target] (target_node) at (T) {};
    \node[anchor=north, font=\scriptsize\bfseries] at ([yshift=-3pt]T) {$\mathbf{p}_u^{\mathrm{target}}$};
    
    \node[selected] (n1_node) at (N1) {};
    \node[selected] (n2_node) at (N2) {};
    \node[selected] at (N3) {};
    \node[selected] at (N4) {};
    \node[selected] at (N5) {};
    
    \node[circle, inner sep=0pt, minimum size=10pt, fill=white] (C_node) at (C) {$\oplus$};
    \node[anchor=north west, font=\scriptsize] at ([xshift=4pt, yshift=2pt]C) {$\mathbf{p}_u^{\mathrm{csp}} = \frac{1}{K} \sum \mathbf{p}_v^{\mathrm{source}}$};
    
    \node[callout, anchor=south] at ([yshift=8pt]T) {\textit{emotional intensity, moral}\\\textit{ambiguity, narrative depth}};
    \node[callout, anchor=south] at ([yshift=8pt]N2) {\textit{atmospheric immersion,}\\\textit{tension-seeking, high agency}};

\end{scope}

\node[target] at (-6.1, -2.8) {}; 
\node[anchor=west, font=\scriptsize] at (-5.9, -2.8) {Query User $u$};

\node[selected] at (-3.6, -2.8) {}; 
\node[anchor=west, font=\scriptsize] at (-3.4, -2.8) {Retrieved Neighbors ($\mathcal{N}_u$)};

\node at (0.6, -2.8) {\scriptsize $\oplus$}; 
\node[anchor=west, font=\scriptsize] at (0.8, -2.8) {Community Source Persona ($\mathbf{p}_u^{\mathrm{csp}}$)};

\end{tikzpicture}
\caption{Visualization of the Community Source Persona aggregation within the shared semantic space. The target user explicitly functions as the query vector to retrieve the $K$ nearest source-domain neighbors via cosine similarity. These proximal representations are mean-pooled to compute the centroid ($\mathbf{p}_u^{\mathrm{csp}}$), which is subsequently transferred to the downstream architecture as the cross-domain persona signal.}
\label{fig:knn_aggregation}
\end{figure}

The resulting CSP captures the common behavioral patterns shared by the retrieved source-domain neighbors. It therefore provides a macroscopic semantic persona about the target user, derived from behaviorally similar users in the source domain rather than from direct identity matching.

The community size $K$ acts as a regularization parameter, independently optimized for each recommendation algorithm and specific source--target domain pairing. A small $K$ may overemphasize idiosyncratic traits of individual source users, whereas an excessively large $K$ may over-smooth the representation into a generic source-domain average. Thus, $K$ controls the tradeoff between specificity and robustness in the transferred semantic signal.

In addition to the CSP, SPHERE can also retain the target user's own persona embedding, $\mathbf{p}_u^{\mathrm{target}}$, as a User Target Persona (UTP). While the CSP provides a cross-domain semantic persona derived from behaviorally similar source users, the UTP provides an in-domain semantic signal derived directly from the target user's own interaction history. We therefore evaluate whether recommendation performance improves when the model relies on the CSP alone, the UTP alone, or a hybrid configuration that jointly incorporates both signals. This hybrid configuration allows the model to learn how the in-domain behavioral description of the target user complements, reinforces, or diverges from the aggregated source-domain behavioral persona, and whether this combination improves recommendation performance in the target recommender system.

\subsection{Semantic Persona Integration and Late-Fusion}\label{sec:architecture}

\begin{figure}[htbp]
\centering
\begin{tikzpicture}[
    >=Latex,
    box/.style={rectangle, draw=black, thick, align=center, rounded corners, minimum height=0.7cm, font=\footnotesize, inner xsep=8pt},
    col_box/.style={box, fill=cyan!10},
    sem_box/.style={box, fill=orange!15},
    fuse_box/.style={box, fill=yellow!15},
    loss_box/.style={box, fill=green!15},
    circ/.style={circle, draw=black, thick, fill=green!15, minimum size=0.9cm, inner sep=0pt, font=\footnotesize\bfseries},
    thick_arrow/.style={draw, line width=1.1pt, ->, >=Latex},
    panel_bg/.style={draw=black, thick, dashed, rounded corners=8pt}
]

\node[col_box] (backbone) at (-4.0, 0) {Backbone Model};

\node[col_box] (user_emb) at (-5.7, 1.3) {User Embedding ($\mathbf{z}_u$)};
\node[col_box] (item_col) at (-2.3, 1.3) {Item Embedding ($\mathbf{z}_i$)};

\draw[thick_arrow] (backbone.north) -- (user_emb.south);
\draw[thick_arrow] (backbone.north) -- (item_col.south);

\node[col_box] (dense1_col) at (-4.0, 2.7) {Dense Layer 1};
\draw[thick_arrow] (user_emb.north) -- (dense1_col.south);
\draw[thick_arrow] (item_col.north) -- (dense1_col.south);

\node (dots_col) at (-4.0, 3.5) {\Large $\vdots$};
\node[anchor=east, font=\scriptsize] at (-4.2, 3.5) {LN + GELU + Drop};

\node[col_box] (denseL_col) at (-4.0, 4.3) {Dense Layer L};
\draw[thick_arrow] (dense1_col.north) -- (dots_col.south);
\draw[thick_arrow] (dots_col.north) -- (denseL_col.south);

\node[sem_box] (csp) at (4.0, 0) {Community Source Persona ($\mathbf{p}_u^{\mathrm{csp}}$)};
\node[sem_box] (proj_csp) at (2.3, 1.3) {Non-Linear Projection};
\node[sem_box] (item_sem) at (5.7, 1.3) {Item Embedding ($\mathbf{e}_i^{\mathrm{sem}}$)};

\draw[thick_arrow] (csp.north) -- (proj_csp.south);

\node[sem_box] (dense1_sem) at (4.0, 2.7) {Dense Layer 1};
\draw[thick_arrow] (proj_csp.north) -- (dense1_sem.south);
\draw[thick_arrow] (item_sem.north) -- (dense1_sem.south);

\node (dots_sem) at (4.0, 3.5) {\Large $\vdots$};
\node[anchor=west, font=\scriptsize] at (4.2, 3.5) {LN + GELU + Drop};

\node[sem_box] (denseL_sem) at (4.0, 4.3) {Dense Layer L};
\draw[thick_arrow] (dense1_sem.north) -- (dots_sem.south);
\draw[thick_arrow] (dots_sem.north) -- (denseL_sem.south);

\node[fuse_box, text width=5.3cm] (fusion) at (0, 5.8) {Fusion Gate ($\alpha_{u,i}$) \& Residual Addition\\ $\mathbf{h}_{u,i}^{\mathrm{col}} + \alpha_{u,i} \cdot \mathbf{h}_{u,i}^{\mathrm{sem}}$};

\draw[thick_arrow] (denseL_col.north) -- node[above left, font=\footnotesize\bfseries, xshift=-0.5cm] {$\mathbf{h}_{u,i}^{\mathrm{col}}$} ([xshift=-1.5cm]fusion.south);
\draw[thick_arrow] (denseL_sem.north) -- node[above right, font=\footnotesize\bfseries, xshift=0.5cm] {$\mathbf{h}_{u,i}^{\mathrm{sem}}$} ([xshift=1.5cm]fusion.south);

\node[fuse_box] (norm) at (0, 7.4) {Layer Normalization};
\draw[thick_arrow] (fusion.north) -- (norm.south);

\node[fuse_box] (pred) at (0, 8.8) {Prediction Head};
\draw[thick_arrow] (norm.north) -- (pred.south);

\node[circ] (y_hat) at (-1.5, 10.2) {$\hat{y}_{u,i}$};
\node[circ] (y) at (1.5, 10.2) {$y_{u,i}$};
\draw[thick_arrow] (pred.north) -- (y_hat.south);

\node[box, fill=white, draw=gray, dashed, font=\scriptsize, inner ysep=4pt] (gt) at (4.0, 10.2) {Ground Truth};
\draw[thick_arrow, dashed, draw=gray] (gt.west) -- (y.east);

\node[loss_box] (loss) at (0, 11.6) {Contrastive Loss Function};
\draw[thick_arrow] (y_hat.north) -- ([xshift=-1cm]loss.south);
\draw[thick_arrow] (y.north) -- ([xshift=1cm]loss.south);

\begin{scope}[on background layer]
    \path (-7.4, -1.2) coordinate (L_bl);
    \path (-0.2, 4.8) coordinate (L_tr);
    \node[panel_bg, fill=gray!4, fit=(L_bl) (L_tr)] (panel_col) {};
    \node[anchor=north, font=\sffamily\bfseries\footnotesize, yshift=0.7cm] at (panel_col.south) {Collaborative Tower};

    \path (0.2, -1.2) coordinate (R_bl);
    \path (7.4, 4.8) coordinate (R_tr);
    \node[panel_bg, fill=gray!8, fit=(R_bl) (R_tr)] (panel_sem) {};
    \node[anchor=north, font=\sffamily\bfseries\footnotesize, yshift=0.7cm] at (panel_sem.south) {Semantic Personas Tower};
\end{scope}

\end{tikzpicture}
\caption{SPHERE dual-tower architecture for the late-fusion of semantic personas and collaborative filtering representations.}
\label{fig:sphere_architecture}
\end{figure}
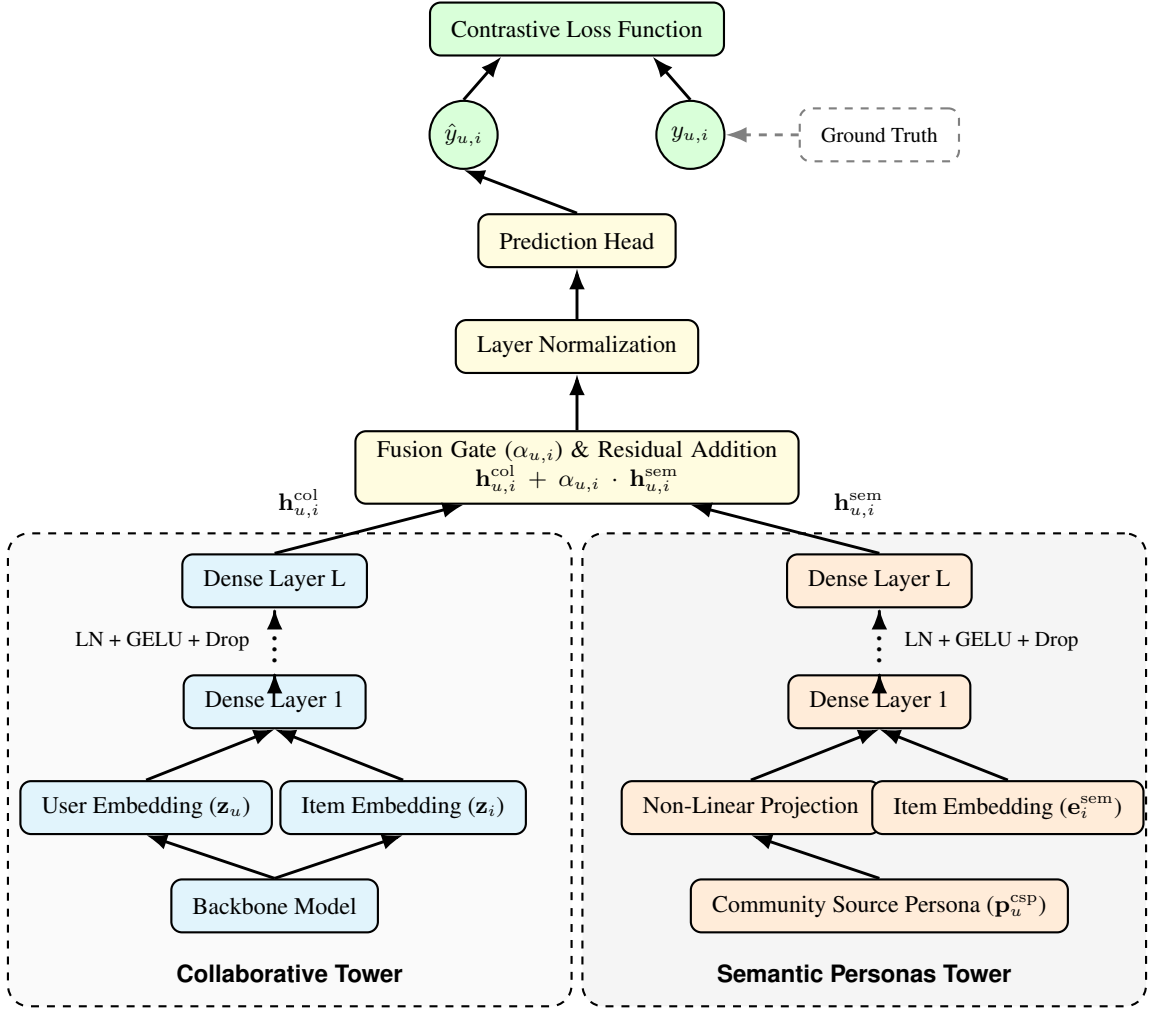

Having established how semantic personas are constructed, we now describe how they are incorporated into the SPHERE recommendation architecture, illustrated in Figure~\ref{fig:sphere_architecture}. SPHERE is designed as a recommender-system design artifact that enables knowledge transfer across strictly disjoint domains. Rather than attempting to align users or items directly, the architecture introduces semantic personas as an intermediate behavioral representation through which source-domain knowledge can inform target-domain recommendation.

The architecture integrates two complementary sources of information. The first is the target-domain collaborative signal, which reflects the structural patterns learned from user-item interactions in the target recommender system. The second is the semantic behavioral signal, which captures persona-based knowledge derived from behaviorally similar users in the source domain. This design allows SPHERE to examine whether behavioral abstractions extracted from an external source domain can complement the target-domain interaction data and improve ranking performance.

The remainder of this section presents the architecture in three stages: collaborative encoding, semantic persona encoding, and dynamic fusion. Together, these components operationalize the core design principle of SPHERE: using shared behavioral representations to transfer recommendation knowledge when conventional identity- or item-based links across domains are unavailable.

\subsubsection{Collaborative Encoding}\label{sec:collab_encoding}\mbox{}\newline
The primary Collaborative tower focuses exclusively on extracting structural, collaborative signals from the target domain interaction history. This architecture is designed to integrate flexibly with various collaborative backbone models, which are trained jointly end-to-end with the broader framework rather than relying on independent, pre-trained weights.

For a given user $u \in \mathcal{U}_T$ and candidate item $i \in \mathcal{I}_T$, we extract their respective dense latent representations, denoted as $\mathbf{z}_u \in \mathbb{R}^d$ and $\mathbf{z}_i \in \mathbb{R}^d$, directly from the collaborative backbone immediately prior to its native scoring operation (such as the final inner product or terminal concatenation).

To project these base representations into a space optimized for cross-domain fusion, the vectors are concatenated and processed through a shared Multi-Layer Perceptron ($\mathrm{MLP}_{\mathrm{col}}$). Crucially, we apply this exact same MLP projection to the terminal end of all isolated baseline models evaluated in our study. This serves a dual purpose: it aligns the structural dimensional space with the semantic tower prior to gating, and it ensures a strictly fair, parameter-matched evaluation. By standardizing the parametric capacity across all architectures, we mathematically guarantee that any observed performance lift is strictly attributable to the semantic persona injection rather than an increase in network depth. The forward pass produces a compact structural interaction vector:
\begin{equation}\label{eq:collab_vector}
\mathbf{h}_{u,i}^{\mathrm{col}} = \mathrm{MLP}_{\mathrm{col}}([\mathbf{z}_u \Vert \mathbf{z}_i])
\end{equation}
where $\Vert$ denotes vector concatenation. The shared network $\mathrm{MLP}_{\mathrm{col}}$ comprises $L$ hidden layers, incorporating layer normalization, non-linear activation functions (e.g., GELU), and dropout regularization to ensure the model learns robust structural features and mitigates overfitting on the sparse interaction matrix.

The resulting output $\mathbf{h}_{u,i}^{\mathrm{col}} \in \mathbb{R}^{d'}$ encapsulates the learned structural interaction between the user's collaborative profile and the specific candidate item. This joint vector forms the primary structural input for the subsequent semantic fusion phase.

\subsubsection{Semantic Encoding}\label{sec:semantic_encoding}\mbox{}\newline
Parallel to the collaborative encoding pipeline, the semantic tower processes the macroscopic behavioral persona retrieved from the source domain. The primary user input to this module is the pre-computed Community Source Persona embedding, denoted as $\mathbf{p}_u^{\mathrm{csp}}$, which encapsulates the transferred behavioral and thematic traits.

Because this dense textual vector originates from a distinct linguistic feature space, we first project it into the collaborative dimensionality $d$ using a dedicated non-linear projection layer:
\begin{equation}\label{eq:csp_projection}
\mathbf{z}_u^{\mathrm{csp}} = \mathrm{GELU}(\mathbf{W}_{\mathrm{proj}} \mathbf{p}_u^{\mathrm{csp}} + \mathbf{b}_{\mathrm{proj}})
\end{equation}
where $\mathbf{W}_{\mathrm{proj}}$ and $\mathbf{b}_{\mathrm{proj}}$ are trainable parameters, and the output is normalized to ensure representational stability.

Crucially, to evaluate the behavioral compatibility between the user's semantic persona and the candidate item, the semantic tower maintains an independent item embedding matrix. For a candidate item $i \in \mathcal{I}_T$, we extract a dedicated semantic item representation $\mathbf{e}_i^{\mathrm{sem}} \in \mathbb{R}^d$. This distinct parameterization prevents representation collapse, allowing the item to maintain a separate structural identity in the collaborative tower and a behavioral identity in the semantic tower.

The projected persona and the semantic item representation are concatenated and processed through a dedicated Multi-Layer Perceptron ($\mathrm{MLP}_{\mathrm{sem}}$) to capture complex behavioral interactions:
\begin{equation}\label{eq:sem_vector}
\mathbf{h}_{u,i}^{\mathrm{sem}} = \mathrm{MLP}_{\mathrm{sem}}([\mathbf{z}_u^{\mathrm{csp}} \Vert \mathbf{e}_i^{\mathrm{sem}}])
\end{equation}
where $\Vert$ denotes vector concatenation. Similar to the structural tower, $\mathrm{MLP}_{\mathrm{sem}}$ consists of $L$ hidden layers with layer normalization, GELU activations, and dropout regularization.

The resulting output is a joint semantic interaction vector $\mathbf{h}_{u,i}^{\mathrm{sem}} \in \mathbb{R}^{d'}$. By architecting this tower to process user-item pairs jointly, the framework ensures that the semantic representation $\mathbf{h}_{u,i}^{\mathrm{sem}}$ is aligned in dimensionality and function with the structural representation $\mathbf{h}_{u,i}^{\mathrm{col}}$, preparing both for the dynamic late-fusion mechanism.

\subsubsection{Dynamic Late-Fusion and Scoring}\label{sec:fusion}\mbox{}\newline
Having extracted the structural interaction vector $\mathbf{h}_{u,i}^{\mathrm{col}}$ and the behavioral interaction vector $\mathbf{h}_{u,i}^{\mathrm{sem}}$, the final architectural stage dynamically integrates these distinct modalities. Because the reliability of the collaborative signal varies drastically depending on a user's target domain sparsity, SPHERE avoids static combination strategies (such as simple addition or fixed-weight concatenation). Instead, it employs an instance-aware gating mechanism to modulate the influence of the semantic persona. We define a learnable fusion gate that computes a scalar modulation weight $\alpha_{u,i} \in (-1, 1)$ for each user-item pair. This weight is determined by jointly evaluating the structural and semantic representations:
\begin{equation}\label{eq:gate}
\alpha_{u,i} = \tanh(\mathbf{W}_{\mathrm{gate}} [\mathbf{h}_{u,i}^{\mathrm{col}} \Vert \mathbf{h}_{u,i}^{\mathrm{sem}}])
\end{equation}
where $\mathbf{W}_{\mathrm{gate}} \in \mathbb{R}^{1 \times 2d'}$ is a bias-free linear transformation. The selection of the hyperbolic tangent ($\tanh$) activation function is deliberate, diverging from conventional gating mechanisms that rely on softmax to produce strictly positive scaling factors, allowing the network to amplify the semantic persona when collaborative data is sparse ($\alpha_{u,i} > 0$), while actively down-scaling or 'moving away' from less relevant semantic contributions ($\alpha_{u,i} < 0$) \citep{hadad2025x}. 

The semantic interaction vector is then scaled by $\alpha_{u,i}$ and applied as a residual update to the primary structural vector. To ensure the feature distributions remain stable prior to final scoring, this composite representation undergoes layer normalization:
\begin{equation}\label{eq:fused_vector}
\mathbf{h}_{u,i}^{\mathrm{fused}} = \mathrm{LayerNorm}(\mathbf{h}_{u,i}^{\mathrm{col}} + \alpha_{u,i} \mathbf{h}_{u,i}^{\mathrm{sem}})
\end{equation}
Finally, the fused representation is projected into a singular, unnormalized prediction logit $\hat{y}_{u,i}$ via a terminal linear scoring head:
\begin{equation}\label{eq:logit}
\hat{y}_{u,i} = \mathbf{W}_{\mathrm{head}} \mathbf{h}_{u,i}^{\mathrm{fused}} + \mathbf{b}_{\mathrm{head}}
\end{equation}
Before this logit is passed to the listwise contrastive loss objective, it is divided by a learnable temperature parameter $\tau$. To maintain gradient stability and prevent the temperature from collapsing or exploding during training, $\tau$ is strictly bounded within an empirically defined operational range:
\begin{equation}\label{eq:scaled_score}
s_{u,i} = \frac{\hat{y}_{u,i}}{\operatorname{clamp}(\tau, \tau_{\min}, \tau_{\max})}
\end{equation}
where $s_{u,i}$ represents the final temperature-scaled affinity score used to calculate the log-probabilities across the sampled candidate items.

\subsubsection{Training Objective}\label{sec:training}\mbox{}\newline
The entire dual-tower architecture, including the collaborative
backbone embeddings and the semantic projection layers, is trained
jointly end-to-end using a listwise contrastive objective. For each
user $u \in \mathcal{U}_T$ interacting with a true target item
$i^+ \in \mathcal{H}_u$, we randomly sample a set of $N$ unobserved
negative items $\mathcal{I}^-$. The model is trained to maximize
the log-probability of the positive interaction against these
candidates using a temperature-scaled cross-entropy loss (InfoNCE):
\begin{equation}\label{eq:loss}
\mathcal{L} = -\sum_{u \in \mathcal{U}_T} \log
\frac{\exp(s_{u,i^+})}{\exp(s_{u,i^+}) + \sum_{j \in \mathcal{I}^-}
\exp(s_{u,j})}
\end{equation}
All optimization details, including learning rate schedules and
regularization, are reported in
Section~\ref{sec:implementation}.

\section{Empirical Evaluation}\label{sec:Evaluation}

\subsection{Datasets and Pre-processing}\label{sec:datasets}
We evaluate the SPHERE framework across three widely adopted real-world datasets: Amazon Books, Goodreads, and Steam \citep{hou2023amazon_dataset, wan2018goodreads_dataset, kang2018steam_dataset}. These domains were deliberately selected to encompass both semantically similar cross-domain scenarios (e.g., Goodreads and Amazon Books) and broadly-relevant cross-domain scenarios (e.g., Steam and Amazon Books). In our experiments, we systematically evaluate all six possible directional target-source pairings among these three domains.

To ensure the reliability of the collaborative signal whilst mitigating the disproportionate influence of highly active power-users, we constrained user profiles to a strict window of $5$ to $15$ interactions. To facilitate computationally tractable persona generation without compromising necessary structural representation, we then extracted a random sample of up to $10,000$ eligible users per domain. 

For evaluation, each user's filtered interaction sequence was ordered chronologically and partitioned using a strict temporal Leave-One-Out (LOO) protocol: the final recorded interaction was reserved for the test set, the penultimate for validation, and all preceding interactions constituted the training set. Crucially, to isolate genuine preference signals and guarantee a high-quality ground truth, we imposed domain-specific positivity thresholds on these evaluation targets. Users in the literary domains (Amazon Books, Goodreads) were retained strictly if both their validation and test items achieved an explicit rating of $\ge 4.0$. Conversely, for the gaming domain (Steam), which employs a binary feedback mechanism rather than an ordinal scale, users were retained only if both evaluation items exhibited a positive recommendation indicator ($= 1$). Any user failing to meet these terminal criteria was entirely excluded from the dataset.

The final network statistics for the sampled domains, which remain structurally consistent whether the domain acts as the target or the source in a given pairing, are detailed in Table \ref{tab:dataset_stats}. The calculation of density reveals that Amazon Books represents our sparsest target environment ($0.1501\%$), while Goodreads ($0.2245\%$) and Steam ($0.2456\%$) exhibit moderately denser interaction topologies.

\begin{table}[htbp]
\TABLE
{Dataset statistics. For each domain, alongside the aggregate number of interactions, we report the structural bipartite graph density. All figures are reported subsequent to the application of the rating thresholds and user sampling protocol.\label{tab:dataset_stats}}
{
\begin{tabular}{l c c c c}
\hline\up 
Domain & Users & Items & Interactions & Density \\ \hline\up 
Amazon Books & 9,961 & 3,839 & 57,383 & 0.1501\% \\
Goodreads & 9,962 & 2,794 & 62,496 & 0.2245\% \\
Steam & 9,968 & 2,470 & 60,464 & 0.2456\% \down\\ \hline
\end{tabular}
}
{}
\end{table}

\subsection{Implementation Details}\label{sec:implementation}
The framework was trained end-to-end utilizing the AdamW optimizer, configured with a learning rate of $3 \times 10^{-3}$ and a weight decay of $1 \times 10^{-3}$, values empirically determined via grid search to provide optimal convergence stability middle-ground across all datasets. Crucially, to ensure a strictly fair comparison and to isolate the specific performance impact of the semantic knowledge transfer, all architectural hyperparameters and training dynamics, including learning rate, regularization, and batch sizes, were held strictly constant across both the isolated baseline models and their respective SPHERE-enhanced variants.

All collaborative backbones utilized a latent embedding dimensionality of $d = 128$ for both user and item representations. The structural interaction tower concatenates the user and item vectors $[\mathbf{z}_u \Vert \mathbf{z}_i] \in \mathbb{R}^{256}$, which are then processed through a three-layer MLP ($256 \rightarrow 128 \rightarrow 64$) with GELU activations, layer normalization, and dropout regularization to produce the structural interaction vector $\mathbf{h}_{u,i}^{\mathrm{col}} \in \mathbb{R}^{64}$. In the semantic tower, the $d_{\text{text}} = 256$-dimensional persona embeddings (UTP and CSP) are each linearly projected to $d = 128$ dimensions prior to concatenation with a dedicated semantic item embedding $\mathbf{e}_i^{\mathrm{sem}} \in \mathbb{R}^{128}$, yielding a joint input $\in \mathbb{R}^{256}$. This vector is processed through a three-layer MLP ($256 \rightarrow 128 \rightarrow 64$) to produce the semantic interaction vector $\mathbf{h}_{u,i}^{\mathrm{sem}} \in \mathbb{R}^{64}$. The scalar fusion gate takes the concatenation $[\mathbf{h}_{u,i}^{\mathrm{col}} \| \mathbf{h}_{u,i}^{\mathrm{sem}}] \in \mathbb{R}^{128}$ as input, and the final fused representation is projected to a scalar prediction score via a single linear layer. All weight matrices were initialized using Kaiming Normal initialization to ensure variance-stable forward propagation at initialization.

To manage memory constraints while preserving a sufficiently large batch size to ensure stable gradient estimation for the contrastive objective, we employed Automatic Mixed Precision (AMP) and processed interactions with a training batch size of $2048$. During training, for each observed positive interaction, we uniformly sampled $N = 8$ unobserved negative items. While theoretical contrastive learning bounds generally benefit from massive negative sample sizes, excessive negative sampling in recommendation contexts frequently induces severe 'false negative' collisions, penalizing unobserved items that are actually relevant to the user. Since $4$--$10$ negatives was shown to be the optimal range for most recommendation settings \citep{ding2020simplify}, we selected $N=8$ as a middle-ground empirical threshold to balance contrastive variance reduction against false-negative penalization.

To stabilize the initial learning dynamics and allow the randomly initialized structural embeddings to develop meaningful collaborative representations before the semantic persona exerts significant gradient pressure, we applied a learning rate schedule comprising a $15$-epoch linear warmup phase followed by cosine annealing decay, alongside gradient norm clipping capped at $1.0$. Models were trained for a maximum of $150$ epochs. We implemented an early stopping mechanism that evaluated the validation NDCG@10 every $5$ epochs, terminating the training procedure if no improvement was observed over a patience window of $30$ epochs to prevent overfitting on the sparse target domains.

For the generation of the semantic personas, we utilized the Llama 3.1 8B Instruct model ($\psi_{\mathrm{gen}}$) \citep{grattafiori2024llama}, selected for its strong zero-shot reasoning capabilities and its suitability for localized use. To map these generated textual profiles into the continuous vector space, we employed OpenAI's \texttt{text-embedding-3-small} model ($\phi_{\mathrm{emb}}$). This model was trained using a Matryoshka Representation Learning (MRL) objective \citep{kusupati2022matryoshka}, which structures the embedding dimensions in descending order of informational importance. This property guarantees that any leading prefix of the full $1536$-dimensional output preserves maximal representational fidelity for that specific dimensionality. Accordingly, we isolate the first $d_{\text{text}} = 256$ dimensions as a principled truncation. By maintaining this intermediate capacity rather than executing a direct reduction to $d = 128$, we substantially mitigate the degradation of the persona vectors prior to their projection into the collaborative feature space during the late-fusion stage.

To optimize the construction of the Community Source Persona (CSP), we treated the semantic community size, $K$, as a tunable hyperparameter. During the validation phase of training, we performed a systematic grid search over $K \in \{5, 15, 30, 50\}$ across multiple seeds. The optimal $K$ was selected dynamically to optimize each specific source-target domain pairing on each backbone model, ensuring the macroscopic persona was appropriately calibrated to the semantic distance between the domains. Furthermore, when aggregating the embeddings of these $K$ neighbors, we deliberately employed an unweighted mean rather than a similarity-weighted average. In highly dense, high-dimensional vector spaces, cosine similarity distributions exhibit extreme top-edge concentration. Consequently, the variance in similarity scores among the top $50$ neighbors out of $10,000$ candidates is mathematically marginal.

\subsection{Baseline Models and SPHERE Configurations}\label{sec:baselines}
To comprehensively and rigorously evaluate the efficacy of the SPHERE framework, we benchmark its performance against three fundamentally distinct structural paradigms within the double-vector collaborative filtering family, ensuring our claims are backbone-agnostic. As established in Section \ref{sec:collab_encoding}, we standardize the deep Multi-Layer Perceptron ($\mathrm{MLP}_{\mathrm{col}}$) as the terminal interaction function across all baselines to isolate the specific impact of the semantic persona.

\begin{itemize}
    \item \textbf{Neural Collaborative Filtering (NCF):} Represents the zero-hop paradigm, utilizing static ID lookups to learn non-linear user-item interaction functions without explicit structural propagation.
    \item \textbf{SVD++:} Represents the one-hop asymmetric paradigm. Based on classic SVD++ mechanics, this backbone enriches the explicit user embedding by aggregating the latent representations of all items in the user's interaction history as an implicit feedback signal.
    \item \textbf{LightGCN:} Represents the multi-hop symmetric paradigm, relying on symmetric normalization and neighborhood aggregation to capture high-order structural proximity.
\end{itemize}

Crucially, we evaluate this framework under conditions of absolute zero-overlap, defined strictly as the environment where $\mathcal{U}_S \cap \mathcal{U}_T = \emptyset$ and $\mathcal{I}_S \cap \mathcal{I}_T = \emptyset$. Classical embedding-alignment methods (e.g., EMCDR) mandate explicit structural overlap, while distributional alignment techniques rely on topological isomorphism that is systematically violated by heterogeneous pairings like gaming versus literature. Furthermore, current semantic-driven approaches rely on complex distributed multi-party training protocols. Because SPHERE transfers holistic abstractions entirely independent of matrix dimensionality, benchmarking against overlap-dependent baselines under equivalent architectural assumptions is methodologically incongruent.

To validate our specific architectural contributions, we compare the primary SPHERE model (CSP) against two Ablation Configurations:
\begin{itemize}
    \item \textbf{SPHERE-Intra:} Replaces the Community Source Persona with the User Target Persona, generated exclusively from their sparse target-domain history, to test if gains can also be achieved by supplementing the user with a semantic representation of themselves.
    \item \textbf{SPHERE-Dual:} Jointly integrates both the in-domain UTP and the cross-domain CSP to evaluate whether providing cumulative dual-domain semantics offers supplementary regularization or introduces redundant semantic interference.
\end{itemize}

\subsection{Evaluation Metrics}\label{sec:metrics}
To assess predictive accuracy and ranking quality, we comprehensively evaluated model performance across a spectrum of standard ranking cutoffs ($k \in \{1, 3, 5, 10, 20\}$), calculating both Hit Ratio (HR) and Normalized Discounted Cumulative Gain (NDCG). However, for clarity and conciseness, we report NDCG@10 throughout this paper, as it is the most widely accepted and rigorously position-aware standard for evaluating top-$N$ recommendation tasks in the current literature. Unlike evaluation protocols that rank the target item against a randomly sampled subset of negative items, a practice known to artificially inflate apparent performance metrics, our evaluation operates over the entire unobserved target item vocabulary. This full-ranking approach ensures a rigorously unbiased evaluation of true retrieval capabilities, though it necessarily yields substantially lower absolute metric values than those reported in studies employing sampled evaluation \citep{krichene2020sampled}.

Recognizing that the sparsity of cross-domain recommendation graphs yields distributions that violate strict parametric assumptions, we establish the statistical significance of the performance lifts exclusively via the one-sided paired t-test over the entire user population. Significance across progressive alpha thresholds ($* p < 0.05$, $** p < 0.01$, $*** p < 0.001$) is calculated at the user level to provide a highly robust, directionally appropriate measure of effect size stability. 

\subsection{Overall Performance and Cross-Domain Transfer Efficacy}\label{sec:overall_performance}

\begin{table}[htbp]
\TABLE
{Main Performance Evaluation (NDCG@10; * $p < 0.05$, ** $p < 0.01$, *** $p < 0.001$). Bold values indicate best performance per row.\textsuperscript{\ddag}\label{tab:main_results}}
{
\begin{tabular}{l l l l l}
\hline\up 
Target & Source & Backbone Model& Baseline& SPHERE\\ \hline\up 
\multirow{6}{*}{Amazon Books} & \multirow{3}{*}{Goodreads} & NCF & 0.0274 & \textbf{0.0311 (+13.43\%)**} \\
& & SVD++ & 0.0295 & \textbf{0.0326 (+10.64\%)*} \\
& & LightGCN & 0.0342 & \textbf{0.0376 (+9.77\%)**} \\ \cline{2-5}
& \multirow{3}{*}{Steam} & NCF & 0.0274 & \textbf{0.0288 (+5.12\%)} \\
& & SVD++ & 0.0295 & \textbf{0.0321 (+8.79\%)*} \\
& & LightGCN & 0.0342 & \textbf{0.0369 (+7.73\%)*} \\ \hline\up
\multirow{6}{*}{Steam} & \multirow{3}{*}{Goodreads} & NCF & 0.0229 & \textbf{0.0259 (+12.81\%)*} \\
& & SVD++ & 0.0181 & \textbf{0.0219 (+21.40\%)**} \\
& & LightGCN & 0.0256 & \textbf{0.0269 (+5.44\%)} \\ \cline{2-5}
& \multirow{3}{*}{Amazon Books} & NCF & 0.0229 & \textbf{0.0260 (+13.33\%)*} \\
& & SVD++ & 0.0181 & \textbf{0.0208 (+15.32\%)*} \\
& & LightGCN & 0.0256 & \textbf{0.0282 (+10.36\%)*} \\ \hline\up
\multirow{6}{*}{Goodreads} & \multirow{3}{*}{Amazon Books} & NCF & 0.0882 & \textbf{0.0937 (+6.29\%)**} \\
& & SVD++ & 0.0909 & \textbf{0.0959 (+5.55\%)*} \\
& & LightGCN & 0.0991 & \textbf{0.1033 (+4.26\%)*} \\ \cline{2-5}
& \multirow{3}{*}{Steam} & NCF & 0.0882 & \textbf{0.0941 (+6.68\%)**} \\
& & SVD++ & 0.0909 & \textbf{0.0974 (+7.17\%)**} \\
& & LightGCN & 0.0991 & \textbf{0.1030 (+3.92\%)*} \down\\ \hline
\multicolumn{5}{l}{\footnotesize \textsuperscript{\ddag}Percentage lifts are derived from full-precision tensor evaluations; displayed NDCG scores are} \\
\multicolumn{5}{l}{\footnotesize rounded to four decimal places, which may result in minor fractional discrepancies if manually recalculated.}
\end{tabular}
}
{}
\end{table}

The primary empirical results, presented in Table \ref{tab:main_results}, demonstrate that the SPHERE framework consistently enhances predictive accuracy across all evaluated target domains and architectural topologies. Crucially, these lifts are achieved under conditions of absolute zero-overlap, proving that macroscopic behavioral personas encoded via large language models can successfully align disjoint latent spaces where traditional geometric mappings fail.

A critical observation from this evaluation is that SPHERE extracts its largest relative performance gains precisely where the structural backbone is most vulnerable. On the Steam target, whose baseline models achieve the lowest absolute NDCG scores across all three domains despite comparable bipartite density (Table~\ref{tab:dataset_stats}), the SVD++ backbone registers the highest absolute lift in the entire evaluation: $+21.40\%$ ($p < 0.01$) using the Goodreads source. This confirms that the semantic abstraction provided by the CSP is most impactful when the native collaborative signal is weakest, a condition that, as discussed below, is governed not merely by graph density but by the discriminative quality of the available interactions.

An important nuance emerges when examining the effect of semantic proximity between the source and target domains. When optimising for the Amazon Books target, the semantically proximate source (Goodreads) yields consistently larger improvements than the distant source (Steam) across all three backbones: $+13.43\%$, $+10.64\%$, and $+9.77\%$ versus $+5.12\%$, $+8.79\%$, and $+7.73\%$ for NCF, SVD++, and LightGCN respectively. However, this proximity advantage does not generalise. On the Steam target, the distant Amazon source matches or outperforms Goodreads for two of three backbones (NCF: $+13.33\%$ vs.\ $+12.81\%$; LightGCN: $+10.36\%$ vs.\ $+5.44\%$), and on the Goodreads target the distant Steam source outperforms Amazon for two of three backbones (SVD++: $+7.17\%$ vs.\ $+5.55\%$; NCF: $+6.68\%$ vs.\ $+6.29\%$). These results indicate that semantic proximity between domains is \textit{not} the primary determinant of transfer efficacy. Rather, the consistent pattern across all permutations is that SPHERE's lift magnitude is governed predominantly by the target domain's native structural performance (cf.\ Section~\ref{sec:gate_dynamics}): the weakest baselines (Steam) attract the largest lifts regardless of which source is paired, while the strongest baselines (Goodreads) benefit more modestly from either source. This finding carries practical significance, as it suggests that practitioners need not restrict cross-domain transfer to semantically adjacent platforms.

\subsection{Cutoff-Dependent Performance Analysis}\label{sec:cutoff_analysis}  
\begin{figure}[htbp]     \centering     \includegraphics[page=1, width=\textwidth]{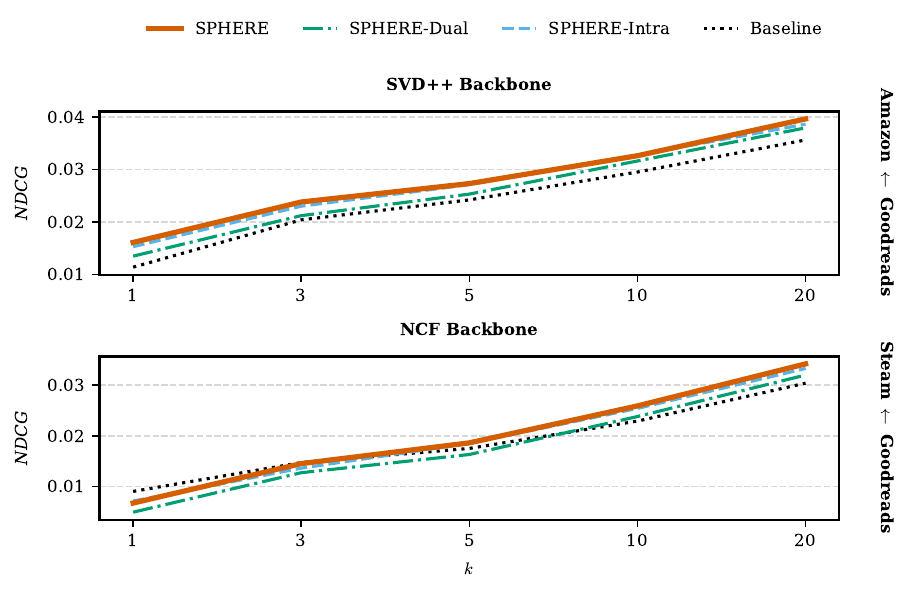}     \caption{NDCG across ranking cutoffs $k \in \{1, 3, 5, 10, 20\}$     for two representative pairings sourced from Goodreads: SVD++ on     Amazon Books (top), illustrating precision concentration at shallow     cutoffs, and NCF on Steam (bottom), illustrating retrieval     elevation at deeper cutoffs. Full results across all domain     permutations, backbones, and Hit Ratio metrics are provided in     Appendix~\ref{app:full_results}.}     \label{fig:ndcg_cutoffs} \end{figure}

To examine how SPHERE's ranking improvements distribute across the recommendation list, Figure~\ref{fig:ndcg_cutoffs} presents NDCG performance across ranking cutoffs $k \in \{1, 3, 5, 10, 20\}$ for two representative pairings that share an identical source domain (Goodreads) but differ in the target baseline's native ranking capacity. All models were optimized exclusively on NDCG@10; the profiles at other cutoffs are therefore emergent properties of the learned representations. These two pairings illustrate the dominant cutoff-dependent patterns observed across all 18 evaluated configurations (full results in Appendix~\ref{app:full_results}).    

A striking divergence reveals two qualitatively distinct mechanisms. On the Amazon target, where the structural baseline achieves moderate performance, SPHERE's lift is sharply concentrated at the top of the list: SVD++ registers $+40.82\%$ at $k{=}1$ ($p < 0.001$), decaying monotonically to $+10.64\%$ at $k{=}10$---a pattern observed in 10 of 18 configurations. This \textit{precision concentration} profile indicates that the backbone already retrieves a viable candidate pool, and the semantic persona acts as a surgical corrector that promotes the relevant item from mid-list to top rank. Conversely, on the Steam target, where baseline NDCG is lowest, the pattern inverts: NCF exhibits a $-25.68\%$ lift at $k{=}1$ but grows to $+12.81\%$ at $k{=}10$ ($p < 0.05$) and $+12.50\%$ at $k{=}20$ ($p < 0.01$)---a \textit{retrieval elevation} pattern observed in 6 of 18 configurations. Here, the collaborative signal alone is insufficient to surface relevant items within the visible window, and SPHERE's contribution is to pull genuinely relevant items \textit{into} the ranking from deep in the unobserved catalog. The gate magnitudes (Table~\ref{tab:fusion_unified}) are consistent with this dichotomy: minimal activation suffices when the baseline is already well-structured, whereas maximal semantic reliance is learned precisely where native ranking capacity is most compromised.

\subsection{Ablation Study: Disentangling Target Density and Semantic Redundancy}\label{sec:ablation}
To rigorously isolate the mechanics driving these performance gains, we deploy an ablation study evaluating the isolated and joint effects of the in-domain User Target Persona (UTP) and the cross-domain Community Source Persona (CSP). This experiment challenges the hypothesis that in-domain semantic summaries are inherently superior to abstract cross-domain representations.

\begin{table}[htbp]
\TABLE
{Impact of Persona Components on NCF Performance (NDCG@10)\textsuperscript{\ddag}\label{tab:ablation_ncf}}
{
\begin{tabular}{l l c c c c}
\hline\up 
Target & Source & Baseline & SPHERE& SPHERE-Dual& SPHERE-Intra\\ \hline\up 
\multirow{2}{*}{Amazon Books} & Goodreads & 0.0274 & 0.0311 (+13.43\%)** & \textbf{0.0313 (+14.22\%)**} & 0.0279 (+1.81\%) \\
       & Steam     & 0.0274 & 0.0288 (+5.12\%) & \textbf{0.0290 (+5.87\%)} & 0.0279 (+1.81\%) \\ \hline\up
\multirow{2}{*}{Steam}  & Goodreads & 0.0229 & \textbf{0.0259 (+12.81\%)*} & 0.0238 (+3.80\%) & 0.0254 (+10.70\%)* \\
       & Amazon Books    & 0.0229 & 0.0260 (+13.33\%)* & \textbf{0.0266 (+16.18\%)**} & 0.0254 (+10.70\%)* \\ \hline\up
\multirow{2}{*}{Goodreads}& Amazon Books  & 0.0882 & 0.0937 (+6.29\%)** & 0.0934 (+5.95\%)* & \textbf{0.0963 (+9.20\%)***} \\
       & Steam     & 0.0882 & 0.0941 (+6.68\%)** & \textbf{0.0977 (+10.84\%)***} & 0.0963 (+9.20\%)*** \down\\ \hline
\end{tabular}
}
{}
\end{table}

\begin{table}[htbp]
\TABLE
{Impact of Persona Components on SVD++ Performance (NDCG@10)\textsuperscript{\ddag}\label{tab:ablation_SVD++}}
{
\begin{tabular}{l l c c c c}
\hline\up 
Target & Source & Baseline & SPHERE& SPHERE-Dual& SPHERE-Intra\\ \hline\up 
\multirow{2}{*}{Amazon Books} & Goodreads & 0.0295 & 0.0326 (+10.64\%)* & 0.0316 (+7.24\%) & \textbf{0.0327 (+10.76\%)*} \\
       & Steam     & 0.0295 & 0.0321 (+8.79\%)* & 0.0324 (+9.87\%)* & \textbf{0.0327 (+10.80\%)*} \\ \hline\up
\multirow{2}{*}{Steam}  & Goodreads & 0.0181 & \textbf{0.0219 (+21.40\%)**} & 0.0202 (+11.51\%) & 0.0197 (+9.25\%) \\
       & Amazon Books    & 0.0181 & 0.0208 (+15.32\%)* & \textbf{0.0210 (+16.40\%)*} & 0.0201 (+11.33\%) \\ \hline\up
\multirow{2}{*}{Goodreads}& Amazon Books  & 0.0909 & \textbf{0.0959 (+5.55\%)*} & 0.0958 (+5.36\%)* & 0.0947 (+4.20\%) \\
       & Steam     & 0.0909 & \textbf{0.0974 (+7.17\%)**} & 0.0957 (+5.25\%)* & 0.0968 (+6.46\%)** \down\\ \hline
\end{tabular}
}
{}
\end{table}

\begin{table}[htbp]
\TABLE
{Impact of Persona Components on LightGCN Performance (NDCG@10)\textsuperscript{\ddag}\label{tab:ablation_lightgcn}}
{
\begin{tabular}{l l c c c c}
\hline\up 
Target & Source & Baseline & SPHERE& SPHERE-Dual& SPHERE-Intra\\ \hline\up 
\multirow{2}{*}{Amazon Books} & Goodreads & 0.0342 & \textbf{0.0376 (+9.77\%)**} & 0.0355 (+3.73\%) & 0.0353 (+3.22\%) \\
       & Steam     & 0.0342 & \textbf{0.0369 (+7.73\%)*} & 0.0361 (+5.44\%) & 0.0356 (+4.06\%) \\ \hline\up
\multirow{2}{*}{Steam}  & Goodreads & 0.0256 & 0.0269 (+5.44\%) & 0.0258 (+1.15\%) & \textbf{0.0280 (+9.42\%)*} \\
       & Amazon Books    & 0.0256 & 0.0282 (+10.36\%)* & \textbf{0.0287 (+12.23\%)*} & 0.0280 (+9.77\%)* \\ \hline\up
\multirow{2}{*}{Goodreads}& Amazon Books  & 0.0991 & \textbf{0.1033 (+4.26\%)*} & 0.1021 (+3.06\%) & 0.1025 (+3.45\%) \\
       & Steam     & 0.0991 & \textbf{0.1030 (+3.92\%)*} & 0.1003 (+1.27\%) & 0.1016 (+2.56\%) \down\\ \hline
\multicolumn{6}{l}{\footnotesize \textsuperscript{\ddag}UTP Only configurations inherit the parameters of their respective optimal CSP source pairing, which accounts for minor fractional} \\
\multicolumn{6}{l}{\footnotesize variance in UTP scores between rows sharing the same target domain.}
\end{tabular}
}
{}
\end{table}

The ablation results across NCF (Table~\ref{tab:ablation_ncf}), SVD++ (Table~\ref{tab:ablation_SVD++}), and LightGCN (Table~\ref{tab:ablation_lightgcn}) reveal a highly sophisticated relationship between structural topology, backbone architecture, and semantic injection. In environments characterized by extreme structural sparsity (e.g., Amazon Books, $0.1501\%$ density), the generated in-domain UTP profiles are inherently under-informed. The linear propagation mechanics of LightGCN are highly sensitive to this semantic noise. In the LightGCN Amazon $\leftarrow$ Goodreads pairing, integrating the target-domain UTP alongside the CSP severely dilutes the efficacy of the model, collapsing the performance lift from $+9.77\%$ (SPHERE) down to $+3.73\%$ (SPHERE-Dual). In these highly sparse regimes, the macroscopic abstraction of the standalone cross-domain CSP consistently provides a much more stable, non-destructive regularization anchor.
Conversely, in the regime where the structural backbone achieves its strongest native performance, Goodreads, which benefits from both moderate density ($0.2245\%$) and high-quality literary metadata, the viability of localized semantic summaries (SPHERE-Intra) depends entirely on the structural capacity of the collaborative backbone. Within the NCF architecture, which utilizes a zero-hop paradigm and inherently lacks a mechanism to explicitly aggregate historical structures, the injection of the UTP is profoundly beneficial. For the NCF Goodreads $\leftarrow$ Steam pairing, the joint SPHERE-Dual configuration yields a massive $+10.84\%$ improvement, and SPHERE-Intra yields $+9.20\%$, both significantly outperforming the standalone SPHERE ($+6.68\%$). Because the native interaction graph supplies rich collaborative signal in this regime, the UTP provides a highly informative, localized behavioral summary that the structurally blind NCF backbone requires to augment its isolated ID representation.
Crucially, this advantage immediately collapses when applied to architectures that possess native mechanisms for structural history modeling. For both SVD++ (which explicitly pools implicit chronological history) and LightGCN (which aggregates multi-hop graph communities), the standalone cross-domain CSP remains the dominant configuration on the Goodreads target. For LightGCN, adding the UTP actually depresses the lift from $+4.26\%$ (SPHERE) down to $+3.06\%$ (SPHERE-Dual). This indicates that when a backbone already explicitly computes the user's localized history via graph or pool routing, injecting a textual summary of that exact same history introduces semantic redundancy that actively interferes with the learned structural representations. 
These findings validate the core architectural contribution of the SPHERE framework: the orthogonal, cross-domain behavioral intelligence provided by the Community Source Persona is universally beneficial across all baseline performance regimes and backbones, whereas in-domain semantic injection is exclusively viable for structurally naive architectures operating in domains where the collaborative backbone already achieves strong native performance.

\subsection{Fusion Gate Dynamics and Semantic Community Analysis}\label{sec:gate_dynamics}
To provide mechanistic insight into how the SPHERE framework leverages semantic knowledge at scale, we analyzed the learned fusion gate activations ($\alpha_{u,i}$) and optimal community configurations across all structural backbones and domain permutations. The gate, defined as $\alpha_{u,i} = \tanh(\mathbf{W}_g [\mathbf{h}_{u,i}^{\mathrm{col}} \| \mathbf{h}_{u,i}^{\mathrm{sem}}])$, modulates the additive contribution of the semantic interaction vector ($\mathbf{h}_{u,i}^{\mathrm{sem}}$) to the structural interaction vector ($\mathbf{h}_{u,i}^{\mathrm{col}}$) via the fused representation $\hat{\mathbf{h}}_{u,i} = \mathbf{h}_{u,i}^{\mathrm{col}} + \alpha_{u,i} \cdot \mathbf{h}_{u,i}^{\mathrm{sem}}$. The unified results are presented in Table~\ref{tab:fusion_unified}.

\begin{table}[htbp]
\TABLE
{Unified Fusion Gate Dynamics and $K$-Neighbor Utilization across architectures. Results report NDCG@10 for the Baseline and SPHERE (CSP Only), relative lift, optimal community size ($K$), and the average learned semantic gate weight ($|\alpha|$).\label{tab:fusion_unified}}
{
\begin{tabular}{l l l c c c c c}
\hline\up
\textbf{Target} & \textbf{Source} & \textbf{Backbone Model}& \textbf{Baseline} & \textbf{SPHERE} & \textbf{Lift (\%)} & \textbf{Opt.\ $K$} & \textbf{Avg.\ $|\alpha|$} \\ \hline\up
\multirow{6}{*}{Amazon Books} & \multirow{3}{*}{Goodreads}
 & SVD++      & 0.0295 & 0.0326 & +10.64\% & 15 & 0.0188 \\
 & & NCF      & 0.0274 & 0.0311 & +13.43\% & 30 & 0.1365 \\
 & & LightGCN & 0.0342 & 0.0376 & +9.77\%  &  5 & 0.1878 \\ \cline{2-8}
 & \multirow{3}{*}{Steam}
 & SVD++      & 0.0295 & 0.0321 & +8.79\%  & 50 & 0.0405 \\
 & & NCF      & 0.0274 & 0.0288 & +5.12\%  &  5 & 0.1766 \\
 & & LightGCN & 0.0342 & 0.0369 & +7.73\%  & 30 & 0.2901 \\ \hline\up
\multirow{6}{*}{Steam} & \multirow{3}{*}{Goodreads}
 & SVD++      & 0.0181 & 0.0219 & +21.40\% & 15 & 0.4561 \\
 & & NCF      & 0.0229 & 0.0259 & +12.81\% & 15 & 0.2185 \\
 & & LightGCN & 0.0256 & 0.0269 & +5.44\%  &  5 & 0.1999 \\ \cline{2-8}
 & \multirow{3}{*}{Amazon Books}
 & SVD++      & 0.0181 & 0.0208 & +15.32\% &  5 & 0.3891 \\
 & & NCF      & 0.0229 & 0.0260 & +13.33\% &  5 & 0.0378 \\
 & & LightGCN & 0.0256 & 0.0282 & +10.36\% & 50 & 0.0240 \\ \hline\up
\multirow{6}{*}{Goodreads} & \multirow{3}{*}{Amazon Books}
 & SVD++      & 0.0909 & 0.0959 & +5.55\%  & 15 & 0.0156 \\
 & & NCF      & 0.0882 & 0.0937 & +6.29\%  & 50 & 0.1254 \\
 & & LightGCN & 0.0991 & 0.1033 & +4.26\%  & 30 & 0.0598 \\ \cline{2-8}
 & \multirow{3}{*}{Steam}
 & SVD++      & 0.0909 & 0.0974 & +7.17\%  & 30 & 0.1162 \\
 & & NCF      & 0.0882 & 0.0941 & +6.68\%  & 30 & 0.1495 \\
 & & LightGCN & 0.0991 & 0.1030 & +3.92\%  & 15 & 0.2319 \down\\ \hline
\end{tabular}
}
{}
\end{table}

\subsubsection{Optimal Community Size ($K$)}
The optimal semantic community size varies substantially across domain pairings and backbones, with no single $K$ universally dominating. This variability validates the grid search protocol over $K \in \{5, 15, 30, 50\}$ and confirms that the appropriate breadth of the Community Source Persona is inherently dependent on the semantic relationship between the paired domains. Notably, smaller communities ($K = 5$) tend to be optimal when the source and target domains share high semantic proximity (e.g., LightGCN on Amazon$\rightarrow$Goodreads), where a tight cluster of highly similar neighbors provides a precise behavioral persona. Conversely, larger communities ($K = 30$--$50$) are frequently selected for more heterogeneous pairings, where broadening the consensus pool compensates for the increased variance in cross-domain behavioral alignment.

\subsubsection{Gate Magnitude ($|\alpha|$) and Performance-Conditioned Modulation}
A striking pattern emerges from the learned gate magnitudes: $|\alpha|$ is systematically governed by the native ranking capacity of the structural backbone on each target domain, rather than by bipartite graph density alone. On the Steam target, where all three backbones achieve their lowest absolute NDCG scores, the gate opens widest: the SVD++ backbone registers $|\alpha| = 0.4561$ for Goodreads$\rightarrow$Steam and $|\alpha| = 0.3891$ for Amazon$\rightarrow$Steam, indicating maximal reliance on the semantic persona where native collaborative signal is weakest. Notably, this occurs despite Steam exhibiting the \textit{highest} bipartite density among the three domains ($0.2456\%$; Table~\ref{tab:dataset_stats}), revealing that graph density is an insufficient predictor of model performance: the inherently higher item-space heterogeneity of interactive entertainment, where user preferences fragment across disparate mechanics, aesthetics, and engagement modalities, produces a noisier collaborative signal per interaction than the comparatively homogeneous literary domains.
SVD++ exhibits the widest dynamic range overall ($|\alpha| \in [0.016, 0.456]$), consistent with its acute sensitivity to the quality of the available collaborative signal: its explicit history aggregation degrades sharply when per-interaction discriminative power is low, necessitating heavy semantic compensation. NCF occupies a narrower middle band ($|\alpha| \in [0.038, 0.219]$), while LightGCN's multi-hop propagation provides implicit structural regularization that stabilizes its gate across performance regimes ($|\alpha| \in [0.024, 0.290]$). Conversely, on Goodreads, where all backbones achieve their strongest native performance, the gate contracts to its lowest magnitudes ($|\alpha|$ as low as $0.016$), confirming that the network learns to inject semantic data only to the extent that the structural backbone's native signal is insufficient.
Despite this contraction, a pronounced \textit{decoupling} between gate magnitude and performance lift reveals the high informational density of the persona signal. The SVD++ backbone on Amazon$\leftarrow$Goodreads achieves $+5.55\%$ lift with $|\alpha| = 0.016$, less than $2\%$ of the structural magnitude, while NCF on Steam$\leftarrow$Amazon delivers $+13.33\%$ lift with $|\alpha| = 0.038$. This decoupling constitutes direct empirical evidence for the semantic orthogonality hypothesis established in the ablation analysis (Section~\ref{sec:ablation}): if the persona signal were collinear with the collaborative embedding, injection at any magnitude would yield negligible lift. The observed pattern, minimal injection along a complementary subspace producing maximal re-ranking effect, confirms that the persona occupies a latent dimension that the structural backbone cannot learn from interaction data alone.

\subsection{Qualitative Analysis: Target User Case Study}\label{sec:case_study}

To concretise the representational impact of semantic knowledge transfer at the individual level, we examine a target user from the Amazon domain augmented by Steam as the source domain. This user was selected because their preference profile and ranking outcomes effectively demonstrate the macro-level trends identified in Sections~\ref{sec:overall_performance}--\ref{sec:gate_dynamics}. The corresponding baseline and SPHERE recommendation lists are detailed in Table~\ref{tab:case_study}.

\textbf{Training Interaction History (8 items):}
\gmark{GenPurple} \textit{Finding Rebecca};
\gmark{GenPurple} \textit{White Rose, Black Forest};
\gmark{GenGray} \textit{Murder One (David Sloane Book 4)};
\gmark{GenGray} \textit{The Jury Master (David Sloane Book 1)};
\gmark{GenGray} \textit{The Short Drop (Gibson Vaughn)};
\gmark{GenPurple} \textit{The Goldfinch: A Novel (Pulitzer Prize for Fiction)};
\gmark{GenBlue} \textit{Being Mortal: Medicine and What Matters in the End};
\gmark{GenPurple} \textit{The Nightingale: A Novel}.

\textbf{Held-Out Test Item:} \gmark{GenRed} \textit{The Butterfly Garden (The Collector Book 1)}

\vspace{2ex}
\noindent\fbox{%
    \parbox{\dimexpr\textwidth-2\fboxsep-2\fboxrule\relax}{%
        \footnotesize \textbf{behavioral Cluster Key:}
        \gmark{GenRed} Psychological Thrillers \& Suspense \quad
        \gmark{GenPurple} Historical \& Literary Fiction \quad
        \gmark{GenGray} Legal Thrillers \& Mystery \quad
        \gmark{GenBlue} Non-Fiction
    }%
}
\vspace{2ex}

\begin{table}[htbp]
\TABLE
{Top-10 Recommendation Lists for the Target User: Baseline vs.\ SPHERE\label{tab:case_study}}
{
\begin{tabular}{p{0.48\textwidth} p{0.48\textwidth}}
\hline\up
\textbf{Isolated NCF Baseline} & \textbf{NCF-SPHERE} \\ \hline\up
1. \gmark{GenGray} \textit{The Racketeer: A Novel} & 1. \gmark{GenRed} \textbf{\textit{The Butterfly Garden (The Collector Book 1)}} \\
2. \gmark{GenGray} \textit{The Temporary Agent (The Agent Book 1)} & 2. \gmark{GenPurple} \textit{Best Kept Secret (Clifton Chronicles Book 3)} \\
3. \gmark{GenGray} \textit{A Criminal Defense (Philadelphia Legal)} & 3. \gmark{GenGray} \textit{Defending Jacob: A Novel} \\
4. \gmark{GenGray} \textit{Damage Control} & 4. \gmark{GenGray} \textit{The Lawyer: A Legal Thriller} \\
5. \gmark{GenGray} \textit{Stay Dead (Elise Sandburg Book 2)} & 5. \gmark{GenRed} \textit{The Girl on the Train: A Novel} \\
6. \gmark{GenGray} \textit{Say You're Sorry (Morgan Dane Book 1)} & 6. \gmark{GenGray} \textit{Before the Fall} \\
7. \gmark{GenGray} \textit{Fatal Decision} & 7. \gmark{GenPurple} \textit{Wolf Hall: A Novel} \\
8. \gmark{GenRed} \textit{The Wolf Road: A Novel} & 8. \gmark{GenGray} \textit{Bodily Harm: A Novel (David Sloane Book 3)} \\
9. \gmark{GenGray} \textit{The Girl in the Ice} & 9. \gmark{GenPurple} \textit{The Art Forger: A Novel} \\
10. \gmark{GenGray} \textit{The Neon Lawyer} & 10. \gmark{GenRed} \textit{The Silent Wife: A Novel} \down\\ \hline
\end{tabular}
}
{}
\end{table}

This user's training history spans three distinct behavioral clusters: historical and literary fiction (\textit{Finding Rebecca}, \textit{White Rose, Black Forest}, \textit{The Goldfinch}, \textit{The Nightingale}), legal thrillers (\textit{Murder One}, \textit{The Jury Master}, \textit{The Short Drop}), and medical non-fiction (\textit{Being Mortal}). The held-out test item, \textit{The Butterfly Garden}, belongs to a fourth cluster---psychological thrillers---that is entirely absent from the training signal. This configuration presents a stringent retrieval challenge: the model must generalize beyond the observed genre distribution to surface an item from an unobserved behavioral mode.

The isolated NCF baseline, relying exclusively on structural co-occurrence patterns within the target collaborative graph, collapses this multi-faceted profile into a near-homogeneous procedural cluster: 9 of its 10 recommended items fall within the legal thriller and mystery category, with a single psychological entry confined to rank 8. The baseline fails to retrieve the test item entirely (NDCG@10 = 0.0000). This outcome reflects a well-documented limitation of collaborative filtering under sparse interaction regimes: the model defaults to the locally densest collaborative community, which in this case corresponds to the legal-thriller subgraph, suppressing the user's literary-fiction preferences whilst marginalizing their psychological-thriller inclinations.

The SPHERE architecture augments this user's representation through the Community Source Persona (CSP), which aggregates behavioral personas from the user's top-$K$ semantic neighbors in the Steam domain. Examining these source personas (Appendix~\ref{app:personas}), a consistent behavioral signature emerges along three of the seven inductively discovered bridge traits (Table~\ref{tab:prompts}): all five neighbors score high on \textsc{Stimulation} (\textit{`high-sensation experiences'}), \textsc{Conflict} (\textit{`tension, risk, and antagonism'}), and \textsc{Investment} (\textit{`long-term, high-friction commitments'}). Notably, the neighbors diverge along the \textsc{Narrative} axis---some emphasize \textit{`structural mechanics and systems'} while others favor \textit{`character-driven storytelling'}---producing an aggregate vector that encodes both psychological intensity and narrative depth rather than collapsing onto a single genre. This composite semantic signal diverges sharply from the formulaic procedural-suspense signal dominant in the target graph's local collaborative community.

When this CSP vector is projected into the collaborative space and modulated by the fusion gate, the resulting re-ranking effects are structurally coherent with the injected semantic signal. First, SPHERE rebalances the genre composition: the proportion of legal thrillers decreases from 9/10 to 4/10, while historical and literary fiction entries (\textit{Best Kept Secret}, \textit{Wolf Hall}, \textit{The Art Forger})---entirely absent from the baseline---appear at ranks 2, 7, and 9, consistent with the CSP's emphasis on \textit{`character-driven storytelling'} and \textit{`emotional resonance'}. Second, the psychological-thriller signal, previously confined to a single low-rank entry in the baseline (rank 8), is elevated to dominate the top of the list: SPHERE surfaces the held-out test item, \textit{The Butterfly Garden}, at rank 1 (NDCG@10 = 1.0000), and promotes two additional psychological thrillers (\textit{The Girl on the Train}, rank 5; \textit{The Silent Wife}, rank 10), consistent with the CSP's strong \textsc{Conflict} and \textsc{Stimulation} activations.

We note that specific ranking changes cannot be attributed to individual persona phrases, as the CSP embedding passes through learned projections and a fusion gate before influencing the final score. However, the systematic correspondence between the dominant trait activations in the source personas and the observed genre redistribution---from procedural homogeneity toward psychological intensity and literary breadth---demonstrates the type of preference disambiguation that purely structural collaborative filtering cannot achieve in sparse interaction regimes.

\section{Conclusion}\label{sec:Conclusion}

In this paper, we introduce SPHERE, a semantic persona-based design artifact for cross-domain recommendation under conditions of complete information siloing, where no users or items are shared across domains. SPHERE addresses a central limitation of persona cross-domain recommender systems: their reliance on identity overlap, shared entities, or structurally comparable graphs. Instead, SPHERE reframes cross-domain transfer as a problem of behavioral semantic alignment. It shifts the basis of alignment from identity to behavior, representing users through LLM-generated, interpretable preference abstractions that can be shared across heterogeneous platforms.

As a design artifact, SPHERE operationalizes this reframing through three core components: an inductive trait discovery procedure that constructs a shared behavioral vocabulary across domains, the generation of structured semantic personas for users, and the aggregation of a Community Source Persona through semantic neighbor retrieval. This persona-based persona is then integrated into existing recommender models, demonstrated with NCF, SVD++, and LightGCN backbones. Our empirical evaluation across Amazon Books, Goodreads, and Steam shows that SPHERE significantly improves recommendation performance across different source-target pairings and model architectures. The strongest gain is observed for Steam as the target domain with Goodreads as the source, where SPHERE improves SVD++ from 0.0181 to 0.0219 NDCG@10, a 21.40\% increase. For Amazon Books, using Goodreads as the source improves NCF, SVD++, and LightGCN by 13.43\%, 10.64\%, and 9.77\%, respectively. These results demonstrate that behavioral personas can provide useful cross-domain personas even when domains are structurally disjoint.

Importantly, the results refine the common intuition that cross-domain transfer is mainly driven by semantic proximity between domains. In several cases, more distant domains provide comparable or stronger gains. For example, when Steam is the target, Amazon Books improves NCF by 13.33\% and LightGCN by 10.36\%, matching or exceeding the gains from Goodreads. This suggests that transfer efficacy depends not only on domain similarity, but also on the structural density and native predictive strength of the target domain.

This study contributes to the literature in three ways. First, it offers a conceptual reframing of cross-domain recommendation from entity-level overlap to interpretable behavioral abstraction. Second, it instantiates this reframing as the SPHERE design artifact, providing actionable design knowledge for building modular, persona-enhanced recommender systems. Third, it shows empirically that semantically distant domains do not necessarily underperform proximate ones, thereby refining prevailing assumptions about when and why cross-domain transfer works.

While SPHERE depends on the quality and stability of LLM-generated personas, future work should examine how different LLMs, prompts, and embedding models affect performance. In addition, the current evaluation is limited to three domains, and future studies should test the framework in additional platform contexts. Finally, although semantic personas reduce the need for direct identity linkage, they raise important questions about privacy, transparency, and fairness in behavioral profiling. Overall, SPHERE demonstrates that behavior-based semantic representations can help overcome information silos and support more interpretable, transferable, and effective personalization across fragmented digital ecosystems.

\clearpage 

\bibliographystyle{plainnat} 
\bibliography{references} 


\appendix

\section{Full Ranking Metric Profiles}\label{app:full_results}

\begin{figure}[htbp]
    \centering
    \includegraphics[page=2, width=\textwidth]{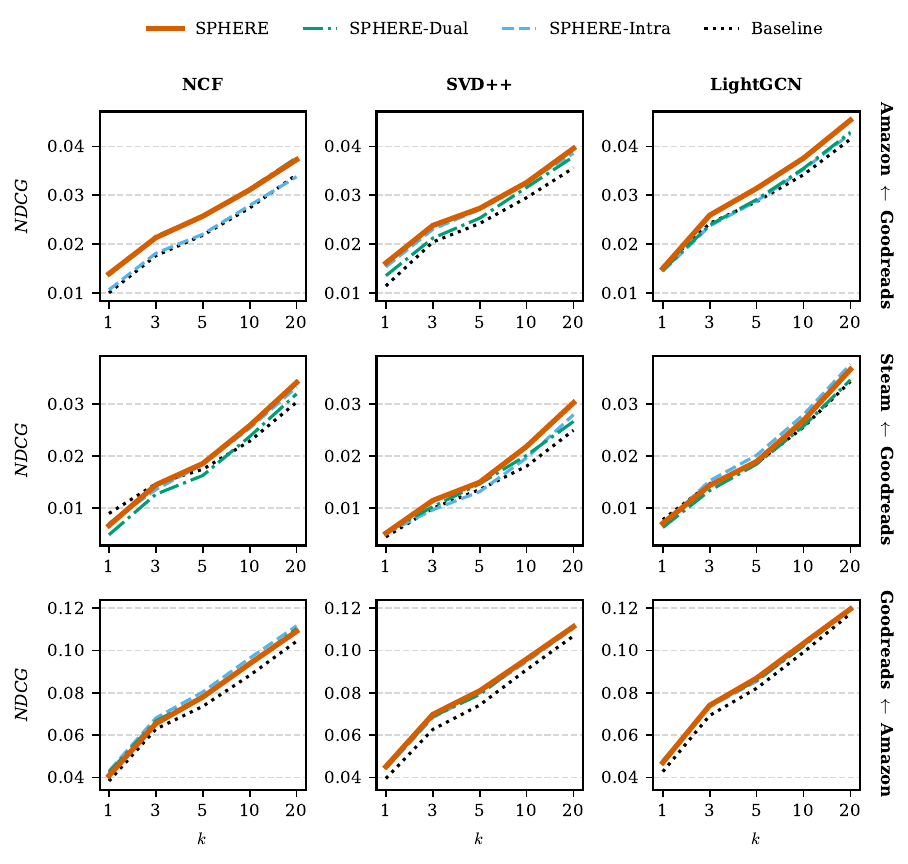}
    \caption{NDCG across ranking cutoffs for all model variants. Rows: Amazon $\leftarrow$ Goodreads, Steam $\leftarrow$ Goodreads, Goodreads $\leftarrow$ Amazon. Columns: NCF, SVD++, LightGCN.}
    \label{fig:app_ndcg_1}
\end{figure}

\begin{figure}[htbp]
    \centering
    \includegraphics[page=3, width=\textwidth]{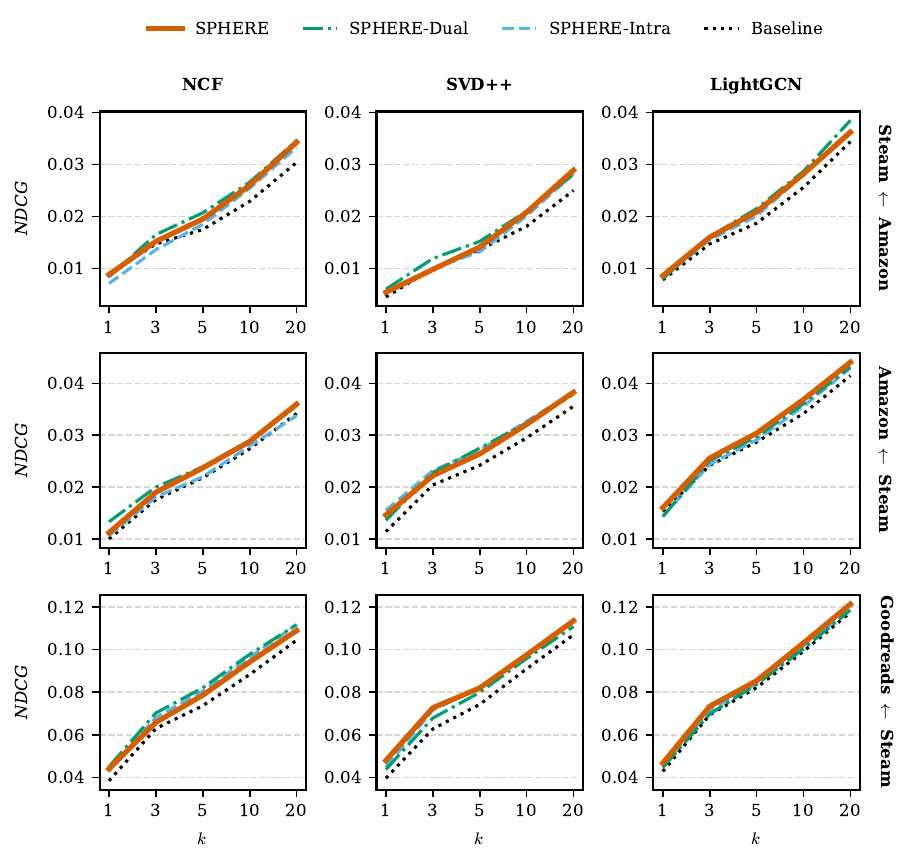}
    \caption{NDCG across ranking cutoffs (continued). Rows: Steam $\leftarrow$ Amazon, Amazon $\leftarrow$ Steam, Goodreads $\leftarrow$ Steam. Columns: NCF, SVD++, LightGCN.}
    \label{fig:app_ndcg_2}
\end{figure}

\begin{figure}[htbp]
    \centering
    \includegraphics[page=4, width=\textwidth]{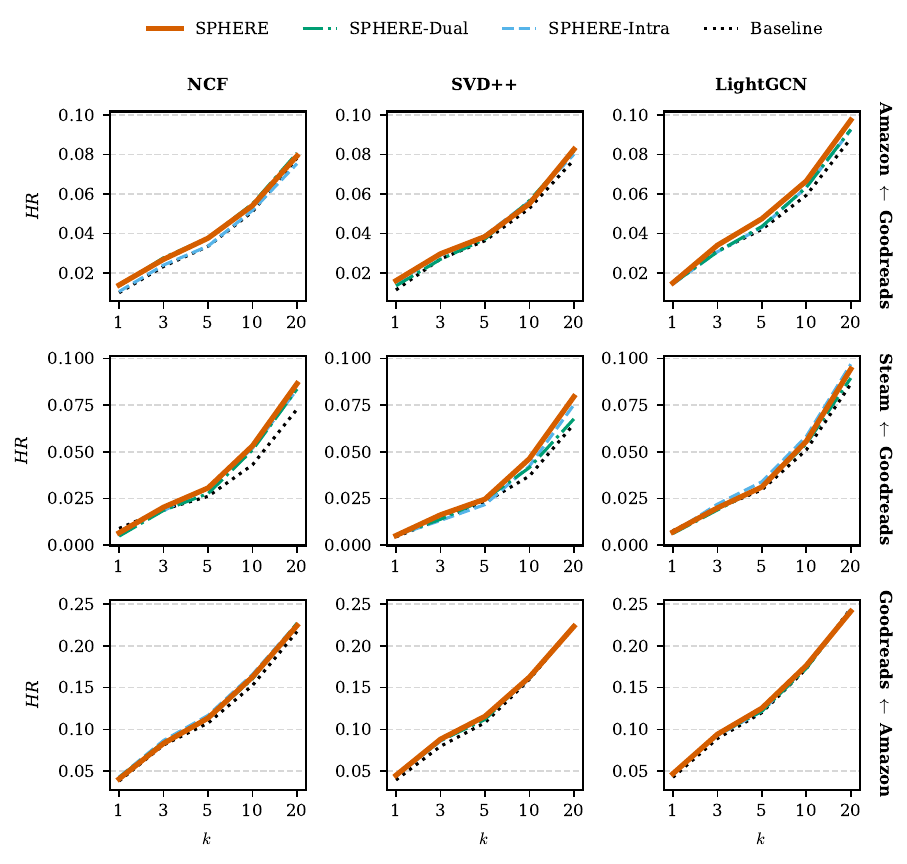}
    \caption{Hit Ratio (HR) across ranking cutoffs for all model variants. Rows: Amazon $\leftarrow$ Goodreads, Steam $\leftarrow$ Goodreads, Goodreads $\leftarrow$ Amazon. Columns: NCF, SVD++, LightGCN.}
    \label{fig:app_hr_1}
\end{figure}

\begin{figure}[htbp]
    \centering
    \includegraphics[page=5, width=\textwidth]{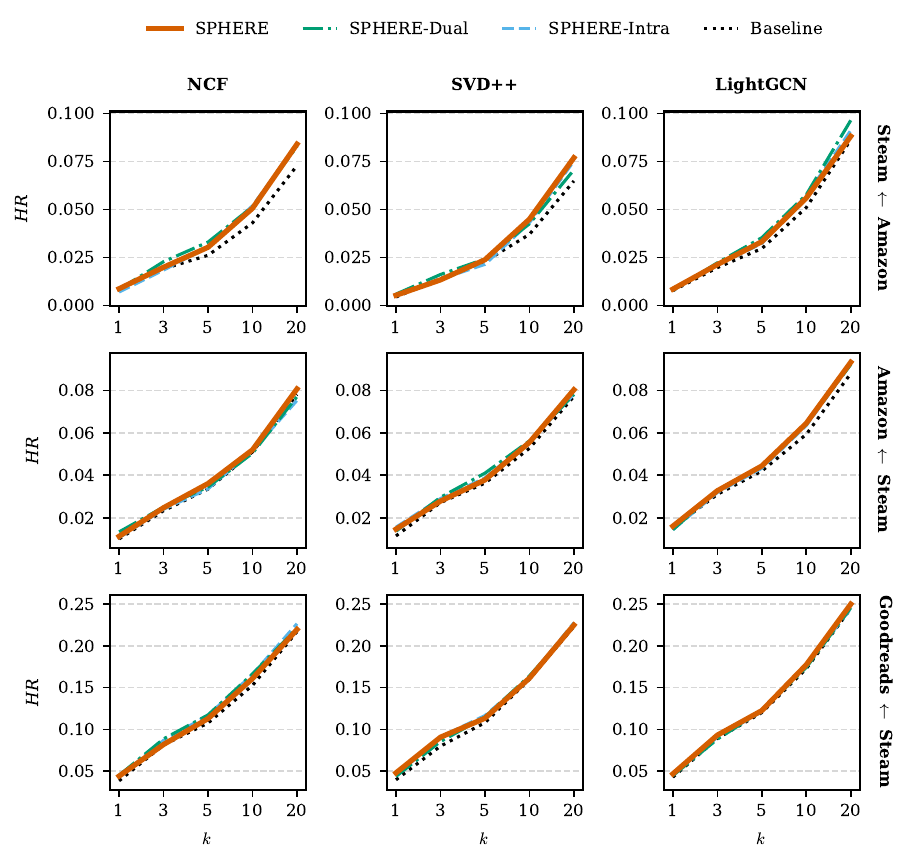}
    \caption{Hit Ratio (HR) across ranking cutoffs (continued). Rows: Steam $\leftarrow$ Amazon, Amazon $\leftarrow$ Steam, Goodreads $\leftarrow$ Steam. Columns: NCF, SVD++, LightGCN.}
    \label{fig:app_hr_2}
\end{figure}

For completeness, Figures~\ref{fig:app_ndcg_1}--\ref{fig:app_hr_2}
report the full NDCG and Hit Ratio (HR) cutoff profiles across all
six domain permutations and three structural backbones. The
cutoff-dependent patterns identified in
Section~\ref{sec:cutoff_analysis} generalize across the remaining
pairings: precision concentration on stronger-baseline targets is
particularly robust under SVD++ and LightGCN, while retrieval
elevation on weaker-baseline targets is most pronounced under NCF and
SVD++. LightGCN's multi-hop propagation partially mitigates baseline
noise on these weaker domains, attenuating the depth-dependent growth
pattern.

\clearpage

\clearpage

\clearpage

\clearpage

\begin{figure}[htbp]
    \centering
    \includegraphics[page=5, width=\textwidth]{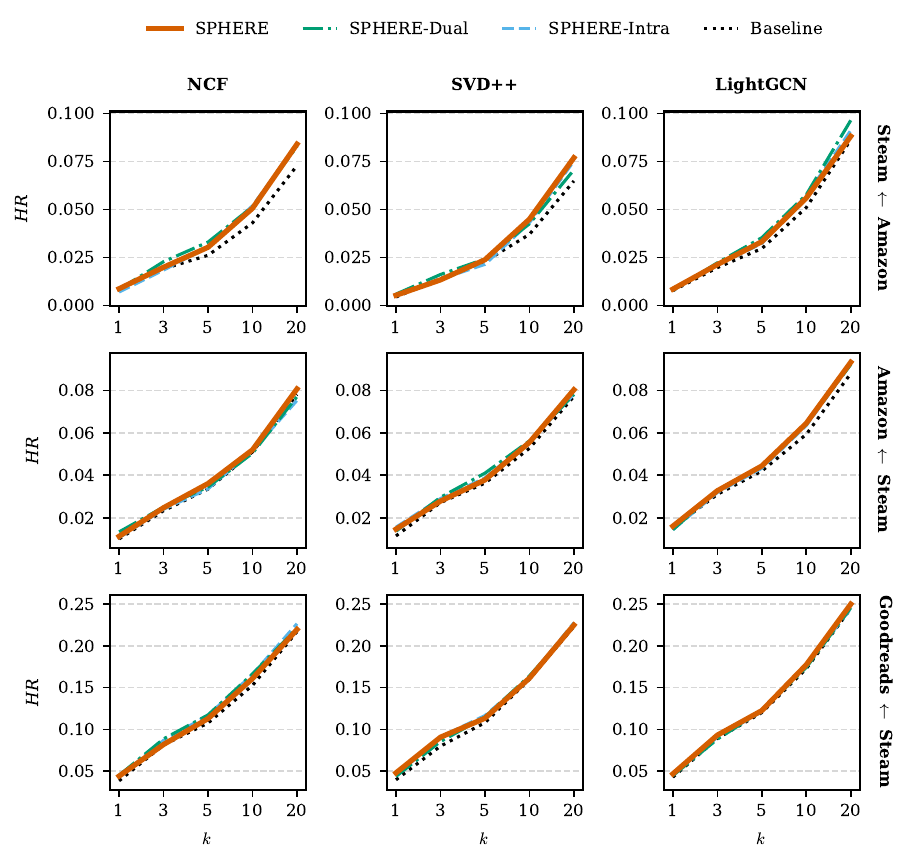}
    \caption{Hit Ratio (HR) across ranking cutoffs (continued).
    Rows: Steam $\leftarrow$ Amazon, Amazon $\leftarrow$ Steam,
    Goodreads $\leftarrow$ Steam. Columns: NCF, SVD++, LightGCN.}
    \label{fig:app_hr_3}
\end{figure}

\clearpage

\section{Semantic Personas: Target User and Source Domain Neighbors}\label{app:personas}

This section first presents the individual User Target Persona (UTP), followed by the raw textual persona representations generated for their $K=5$ nearest semantic neighbors utilized to construct the Community Source Persona (CSP) in Section~\ref{sec:case_study}. As detailed below, the retrieved neighbors from the Steam domain predominantly exhibit an attraction to emotionally charged, high-friction, and tension-heavy interactive environments.

\vspace{2ex}
\noindent\textbf{Target User Persona (UTP)}
\nopagebreak\smallskip\nopagebreak

\noindent\textit{'The user's behavioral profile indicates a preference for high-sensation experiences, often drawn to intense, emotionally charged narratives that explore complex themes and moral ambiguities. This is reflected in their attraction to stories that balance structural mechanics with character-driven storytelling, and their tolerance for long-term, high-friction commitments to characters and their struggles, even in the face of uncertainty and conflict.'}

\vspace{2ex}
\noindent\textbf{Source Neighbor 1} (Sim: 0.9461)
\nopagebreak\smallskip\nopagebreak

\noindent\textit{'The user's behavioral profile indicates a preference for high-sensation experiences, often seeking out challenging and difficult content, while also valuing long-term, high-friction commitments and investments in complex systems and narratives. This is reflected in their attraction to tension, risk, and antagonism, as well as their motivation driven by structural mechanics and systems, rather than character-driven storytelling.'}

\vspace{2ex}
\noindent\textbf{Source Neighbor 2} (Sim: 0.9401)
\nopagebreak\smallskip\nopagebreak

\noindent\textit{'The user's behavioral profile indicates a preference for high-sensation experiences, often seeking out immersive, atmospheric, and emotionally charged narratives that frequently incorporate elements of conflict, tension, and risk. This is balanced by a desire for creative expression and individuality, as well as a need for long-term, high-friction commitments that allow for systemic problem-solving and character-driven storytelling.'}

\vspace{2ex}
\noindent\textbf{Source Neighbor 3} (Sim: 0.9388)
\nopagebreak\smallskip\nopagebreak

\noindent\textit{'The user's behavioral profile indicates a preference for high-sensation experiences with a strong desire for emotional resonance and character-driven storytelling, often accompanied by a need for internal agency and control, as well as a tolerance for long-term, high-friction commitments that allow for complex problem-solving and nuanced narrative exploration. This is also reflected in a willingness to engage with themes of conflict, tension, and antagonism, while also valuing relationships, trust, and emotional intimacy.'}

\vspace{2ex}
\noindent\textbf{Source Neighbor 4} (Sim: 0.9321)
\nopagebreak\smallskip\nopagebreak

\noindent\textit{'The user's behavioral profile indicates a preference for high-sensation experiences, often engaging with narratives that involve tension, risk, and antagonism, and a strong desire for internal agency/control, as evidenced by their attraction to complex, systemic problem-solving and immersive, atmospheric environments. This is further supported by their tolerance for long-term, high-friction commitments and their motivation driven by structural mechanics/systems, rather than character-driven storytelling.'}

\vspace{2ex}
\noindent\textbf{Source Neighbor 5} (Sim: 0.9314)
\nopagebreak\smallskip\nopagebreak

\noindent\textit{'The user's behavioral profile indicates a preference for high-sensation experiences, often seeking out complex narratives with a mix of structural mechanics and character-driven storytelling, and a need to express individuality through creative problem-solving and self-directed agency, while also being drawn to atmospheric, emotional, and cinematic experiences that explore themes of conflict, risk, and antagonism. This profile suggests a tolerance for long-term, high-friction commitments and a willingness to engage with challenging, often violent, content, but also a desire for deep, character-driven connections and meaningful narrative resolutions.'}


\end{document}